\documentclass[aps,prd,notitlepage,showpacs,nofootinbib,superscriptaddress]{revtex4-1}

\usepackage{graphicx}
\usepackage[utf8]{inputenc} 

\usepackage{amsmath}
\usepackage{amssymb}
\usepackage{braket}
\usepackage{float}
\usepackage{comment}
\usepackage{slashed}
\usepackage[normalem]{ulem}

\usepackage{caption}
\usepackage{subcaption}
\usepackage{amsmath}
\usepackage{amssymb}

\usepackage[usenames,dvipsnames]{color}
\usepackage[colorinlistoftodos]{todonotes}
\usepackage[colorlinks=true,citecolor=darkred,urlcolor=darkred, pdfborder={0 0 0}]{hyperref}
\usepackage[normalem]{ulem}

\definecolor{darkred}{rgb}{0.6,0,0}
\definecolor{darkbrown}{rgb}{0.39,0.26,0.12}
\usepackage[colorinlistoftodos]{todonotes}

\definecolor{linkcolor}{rgb}{0,0,0.5}
 
\newcommand {\ignore}[1]{}

\def \znbb {$\rm 0\nu\beta\beta$}

\def\gsim{\raise0.3ex\hbox{$\;>$\kern-0.75em\raise-1.1ex\hbox{$\sim\;$}}}
\def\lsim{\raise0.3ex\hbox{$\;<$\kern-0.75em\raise-1.1ex\hbox{$\sim\;$}}}
\def\lfv{lepton flavour violation }

\def\SM{$\mathrm{SU(3)_c \otimes SU(2)_L \otimes U(1)_Y}$ }
\newcommand{\sm}{{Standard Model }}
\def\vb#1{\vbox to #1 pt{}}

\usepackage{soul}

\definecolor{mightnightblue}{RGB}{25,25,112}

\definecolor{brown}{rgb}{0.59, 0.29, 0.0}

\def\lfv{lepton flavour violation }

\def\SM{$\mathrm{SU(3)_c \otimes SU(2)_L \otimes U(1)_Y}$ }
\def\21{$\mathrm{SU(2)_L \otimes U(1)_Y}$}
\def\lfv{lepton flavour violation }

\def\sm{standard model }

\newcommand{\AddrAHEP}{%
  AHEP Group, Institut de F\'{i}sica Corpuscular --
  CSIC/Universitat de Val\`{e}ncia, Parc Cient\'ific de Paterna.\\
 C/ Catedr\'atico Jos\'e Beltr\'an, 2 E-46980 Paterna (Valencia) - SPAIN}

\newcommand{\AddrIISERB}{Department of Physics, Indian Institute of Science Education and Research - Bhopal, \\ 
Bhopal Bypass Road, Bhauri, Bhopal 462066, India}

\begin{document}
\bibliographystyle{unsrt}   

\title{\color{BrickRed} Dark matter as the origin of neutrino mass in the inverse seesaw mechanism}

\author{Sanjoy Mandal}\email{smandal@ific.uv.es}
\affiliation{~\AddrAHEP}

\author{Nicol\'{a}s Rojas}\email{nicolas.rojasro@usm.cl}
\affiliation{Universidad T\'ecnica Federico Santa Mar\'{i}a - Departamento de F\'{i}sica\\
      Casilla 110-V, Avda. Espa\~na 1680, Valpara\'{i}so, Chile}
      
\author{Rahul Srivastava}\email{rahul@iiserb.ac.in}
\affiliation{\AddrIISERB}

\author{Jos\'{e} W. F. Valle} \email{valle@ific.uv.es} 
\affiliation{~\AddrAHEP}

\begin{abstract}
 
  We propose that neutrino masses are ``seeded'' by a dark sector within the inverse seesaw mechanism.
  This way we have a new, ``hidden'', variant of the scotogenic scenario for radiative neutrino masses.
  We discuss both explicit and dynamical lepton number violation.
  In addition to invisible Higgs decays with majoron emission, we discuss in detail the pheneomenolgy of dark matter, as well as the novel features associated to charged lepton flavour violation,
  and neutrino physics. 
\end{abstract}

\maketitle


\section{Introduction}

Two irrefutable evidences of new physics are the existence of neutrino mass~\cite{McDonald:2016ixn,Kajita:2016cak} and dark matter~\cite{Bertone:2004pz}.
Neither can be explained within the framework of Standard Model. There are many extensions of the Standard Model which can induce neutrino masses. 
There are also many extensions of the Standard Model that predict the existence of new electrically neutral fermions and/or scalars that can be made stable
by imposing appropriate symmetries, such as to play the role of dark matter. In most of these extensions \emph{a priori} there is no relation between dark matter and neutrino mass generation. 
Our focus here is to investigate models where neutrino mass generation is intimately connected with dark matter.
One possibility is that dark matter is the mediator of neutrino mass generation~\cite{Ma:2006km,Hirsch:2013ola,Merle:2016scw,Rocha-Moran:2016enp,Restrepo:2019ilz,Avila:2019hhv}. 
One can also combine such radiative ``scotogenic'' scenario with the seesaw paradigm, as in the scoto-seesaw models~\cite{Rojas:2018wym,Barreiros:2020gxu,Mandal:2021yph}.
For example, this can be achieved by combining the simplest (3,1) version of the seesaw mechanism containing a single heavy ``right-handed'' neutrino~\cite{Schechter:1980gr},
with the minimal scotogenic scenario for dark matter~\cite{Ma:2006km}. 
In such hybrid scenario the ``atmospheric scale'' comes from the tree level seesaw mechanism, while the ``solar scale'' is mediated by the radiative exchange of dark states. 
This gives a clear interpretation of the hierarchy in ``atmospheric'' and ``solar'' oscillation lengths, as well as a viable WIMP dark matter candidate.\\[-.3cm] 

Here we propose a simpler way to combine the seesaw and the scotogenic approaches, by making dark matter the seed of neutrino mass generation
within a low-scale seesaw mechanism.
For definiteness we take the inverse seesaw as our template~\cite{Mohapatra:1986bd,Gonzalez-Garcia:1988okv}. 
The interest on such low-scale approach~\cite{Boucenna:2014zba} is not only conceptual, but can also lead to distinctive phenomenological implications.
Not only the dynamics of electroweak breaking~\cite{Joshipura:1992hp} gets modified, 
but also important conceptual features emerge in the flavor and CP sector~\cite{Bernabeu:1987gr,Branco:1989bn,Rius:1989gk}, many of which can be probed in experiment.
It has been proposed that in inverse seesaw schemes, such as ours, the mediators of neutrino mass generation can be searched for at high energy colliders~\cite{Dittmar:1989yg,Gonzalez-Garcia:1990sbd}.
The latest restrictions come for the ATLAS and CMS experiments~\cite{CMS:2018iaf}. Likewise, as originally proposed, the \lfv phenomenology can be probed at high energies and high intensity 
facilities~\cite{Gonzalez-Garcia:1991brm,Ilakovac:1994kj,Deppisch:2004fa,Deppisch:2005zm,DeRomeri:2016gum}. 
The latter feature is also characteristic of the inverse seesaw mechanism. \\[-.3cm] 

Here we examine our proposal that neutrino masses are ``seeded'' by a dark sector within the inverse seesaw mechanism.
We show that the phenomenology of dark matter differs from that of the original scotogenic scenario, since dark matter lives in a ``hidden'' sector, singlet under the \SM gauge symmetry.
Detailed predictions and therefore distinct from those of existing scotogenic scenarios~\cite{Ma:2006km,Hirsch:2013ola,Merle:2016scw,Rocha-Moran:2016enp,Restrepo:2019ilz,Avila:2019hhv}.
We also show how the scheme can be independently probed in a variety of ways in the flavor, collider and neutrino sectors. \\[-.3cm] 

The paper is organized as follows. To start with, in sec.~\ref{sec:explicit-breaking}, we discuss the simplest radiative generation of neutrino mass scheme with explicit breaking of lepton number
proposed in~\cite{Ahriche:2016acx}. 
Building upon on this discussion, in sec.~\ref{sec:dynamical-breaking} we discuss how one can generate neutrino mass through dynamical breaking of lepton number. 
In sec.~\ref{sec:Higgs-physics} we describe the phenomenology of Higgs boson physics and in sec.~\ref{sec:dark-matter} we discuss both the primordial dark matter relic density 
and also the direct detection prospects. 
In sec.~\ref{sec:LFV} we discuss various phenomenological implications for lepton flavour violation, mentioning briefly 
 the prospects for detection of the inverse seesaw neutrino mass mediators at colliders. 
 Finally, in sec.~\ref{sec:neutr-phen} we comment on the neutrino phenomenology \emph{per se} including a possibly detectable neutrinoless double beta decay signal as well as
 deviations from the three-neutrino paradigm in neutrino oscillations.
 We conclude and summarize our main results in sec.~\ref{sec:summary}.

\section{Explicit lepton number breaking} 
\label{sec:explicit-breaking}

Gauge singlet fermions are unconstrained by anomaly requirements, hence we can assume any number of them~\cite{Schechter:1980gr}.
This is used as a key feature of the inverse seesaw mechanism, which extends the standard \SM model with new fermions $\nu^c$, $S$.
The minimum multiplicity generating the two mass splittings required for the interpretation of the neutrino oscillation data is having two $\nu^c$, $S$ pairs.
This can account both for the atmospheric and solar mass scales at tree-level.
Counting also the three active neutrinos, such setup corresponds to a (3,2,2) scheme, in the language of~\cite{Schechter:1980gr}\footnote{In our construction we have also the multiplicity
of dark fermions $f$ in the loop, which we assume to match that of $S$.}.
However, throughout most of this paper we will make the more common, non-minimal, sequential choice of having three $\nu^c$, $S$ pairs, or (3,3,3).\\[-.4cm]

The simplest way to have a ``dark sector'' as origin for neutrino masses in such inverse seesaw mechanism is to postulate the existence
of an extra dark fermion $f$, as well as a new complex dark scalar $\xi$~\cite{Ahriche:2016acx}.
The new particles and their charges are given in Table.~\ref{tab:MatterModel}. 
\begin{table}[!h]
\setlength\tabcolsep{0.25cm}
\centering
\begin{tabular}{| c || c | c | c | c | c || c | c  |}
\hline 
                             &   $L$         &     $e^c$     &    $\nu^c$     &     $S$        &     $f$       &   $\Phi$         & $\xi$        \\
\hline     \hline                                                                                                            
$SU(2)_L\times U(1)_Y$       &  $(2,-1)$     &  $(1,2)$     &  $(1,0)$       &    $(1,0)$     &    $(1,0)$    &   $(2,1)$  &   $(1,0)$    \\
$U(1)_{B-L}$                     &   $-1$        &   $1$        &    $1$         &    $-1$        &     $0$       &    $0$     &  $1$       \\
$\mathcal{Z}_2$              &    $+$        &    $+$        &    $+$         &    $+$         &     $-$       &    $+$     &  $-$          \\
\hline
\end{tabular}
\caption{\label{tab:MatterModel}Fields and their quantum numbers.}
\end{table}

In Table.~\ref{tab:MatterModel} the additional $\mathcal{Z}_2$ symmetry is the “dark parity” responsible for the stability of the dark matter candidate.
The relevant terms in the Yukawa Lagrangian responsible for dark matter and neutrino mass generation are the following:
\begin{eqnarray}
\mathcal{L}^{\text{Yukawa}} &=& -\,Y_{\nu^c} \overline{L} i\tau_2 \Phi^* \nu^c\, -\,Y_\xi \xi f S\,-\,M \nu^c S \,-\, \frac{1}{2}\mathcal{M}_f f f\, + H.c.
\label{eq:Yukawa-Lagrangian} 
\end{eqnarray}
where all Yukawa couplings and mass parameters are $3\times 3$ matrices. The interactions in Eq.\eqref{eq:Yukawa-Lagrangian} are invariant under the \sm gauge group and, 
in addition, under the $\mathcal{Z}_2\times U(1)_{B-L}$ symmetry associated to dark matter and lepton number. \\[-.4cm]

The scalar potential is given by, 
\begin{eqnarray} 
\mathcal{V}_{(s)} & = & -m^2 \Phi^\dagger \Phi \,+ \, \frac{\lambda_\Phi}{2}\left( \Phi^\dagger \Phi\right)^2
\,- \, m_\xi^2 \xi^* \xi \,+\,\frac{\lambda_\xi}{2}\left( \xi^* \xi\right)^2 
\,+ \, \lambda_{\Phi\xi} \left(\Phi^\dagger \Phi \right)\left(\xi^* \xi \right) 
\,+\,\frac{\mu_\xi^2}{4}\left(\xi^2 \, + \, \text{h.c.}\right) 
\label{eq:scapot-explicit}
\end{eqnarray}
 Although, the above potential is $\mathcal{Z}_2$ symmetric and renormalizable, the term $\mu_\xi$ breaks the $U(1)_{B-L}$ symmetry softly. 
This is required to generate the neutrino masses, as it is the origin of lepton number violation by two units, see further discussion later.\\[-.4cm] 

In order to ensure dark matter stability the $\mathcal{Z}_2$ symmetry should remain unbroken.
This means that the $\mathcal{Z}_2$ odd scalar $\xi$ should not acquire a nonzero vacuum expectation value (vev).
As a result, electroweak symmetry breaking is driven simply by the vev of $\Phi$. The fields $\xi$ and $\Phi$ can be written as follows 
\begin{align}
\Phi=
\begin{pmatrix}
\phi^+\\
(v_\Phi+h+i\phi^0)/\sqrt{2}
\end{pmatrix},\,\,
\xi=(\xi_R+i\xi_I)/\sqrt{2}
\end{align}
Exact conservation of the $\mathcal{Z}_{2}$ symmetry forbids the mixing between the Higgs and the $\xi$. 
The Higgs mass is exactly the same as in the standard model, $m_h^2=\lambda_\Phi v_\Phi^2$.   The real and imaginary components of $\xi$ have the following masses
\begin{align}
m_{\xi_R}^2&=m_\xi^2+\frac{1}{2}\left(\lambda_{\Phi\xi}v_\Phi^2+\mu_\xi^2\right)\\
m_{\xi_I}^2&=m_\xi^2+\frac{1}{2}\left(\lambda_{\Phi\xi}v_\Phi^2-\mu_\xi^2\right)
\end{align}
\begin{figure}[htb!]
\centering
\includegraphics[width=0.4\textwidth]{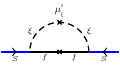}
\caption{``Scotogenic'' origin for the lepton number violating $\mu$-term in the simplest inverse seesaw mechanism.}
\label{fig:scoto-loop-explicit}
\end{figure}
The difference $m_{\xi_R}^2-m_{\xi_I}^2$ depends only on the parameter $\mu_\xi$ which, we will show later, is also responsible for smallness of neutrino masses.
The conservation of the $\mathcal{Z}_2$ symmetry also ensures the stability of the lightest of the two scalar eigenstates $\xi_R$ and $\xi_I$.
As we will show below, this will be a viable dark matter candidate.  \\[-.4cm] 

We now turn to the issue of radiative neutrino mass generation in the case of explicit breaking of lepton number.
The tree-level neutral lepton mass matrix has the following \emph{block} structure in the basis $(\nu, \nu^c, S)$ of our (3,3,3) scheme,
\begin{eqnarray}
\mathcal{M}_{F^0} &=& \left[ \begin{array}{ccc}
                               0 & m_D & 0 \\
                               m_D & 0 & M \\
                               0 & M & 0 
                               \end{array} \right] \,\,
                   \,\,=\,\, \left[ \begin{array}{ccc}
                               0 & Y_{\nu^c}^i \frac{v_\Phi}{\sqrt{2}} & 0 \\
                               Y_{\nu^c}^j \frac{v_\Phi}{\sqrt{2}} & 0 & M \\
                               0 & M & 0 
                               \end{array} \right]. 
       \label{eq:treemass}
\end{eqnarray}
Clearly, the effective light neutrino mass matrix obtained by ``integrating-out'' the heavy singlets is exactly zero, so that at the tree level neutrinos are  
massless~\footnote{One can see this also by using the method of perturbative seesaw diagonalization by blocks developed in~\cite{Schechter:1981cv}.}. 
This is the template for implementing the inverse-seesaw mechanism~\cite{Mohapatra:1986bd,Gonzalez-Garcia:1988okv}. 
Notice that, within the conventional approach, there is a nonzero tree-level $SS$ mass entry, which leads to tree-level neutrino masses, through: 
\begin{align}
\mathcal{M}_\nu=m_D M^{-1}\mu M^{-1T}m_D^T.
  \label{eq:iss}
\end{align}
Small neutrino masses are protected by lepton number symmetry hence natural in t'Hooft's sense.
Here, however, this term arises only as a result of calculable radiative correction indicated in Fig.~\ref{fig:scoto-loop-explicit}. 
Hence neutrinos are massless at the tree approximation, and one has an extra protection for their small masses coming from the loop. 
Indeed, turning on the soft-breaking-term $\frac{\mu_\xi^2}{4} (\xi^2+\text{h.c.})$  in the potential, lepton number gets violated and generates the effective $\mu$ term in Eq.~(\ref{eq:iss})
through the one loop diagram given in Fig.~\ref{fig:scoto-loop-explicit}. The result is 
\begin{align}
\mu&=\frac{Y_\xi Y_\xi}{16\pi^2}\mathcal{M}_f\left(B_0(0,\mathcal{M}_f^2,m_{\xi_R}^2)-B_0(0,\mathcal{M}_f^2,m_{\xi_I}^2)\right)\nonumber \\
&=\frac{Y_\xi Y_\xi}{16\pi^2}\mathcal{M}_f\left( \frac{m_{\xi_R}^2}{m_{\xi_R}^2-\mathcal{M}_{f}^2}\log \left( \frac{m_{\xi_R}^2}{\mathcal{M}_{f}^2}\right) \,-\, \frac{m_{\xi_I}^2}{m_{\xi_I}^2-\mathcal{M}_{f}^2}\log \left( \frac{m_{\xi_I}^2}{\mathcal{M}_{f}^2} \right) \right)
\end{align}
where $B_0(0,\mathcal{M}_f^2,m_{\xi_{R/I}}^2)$ is a Passarino-Veltman function evaluated in the limit of zero external momentum.

\section{Dynamical lepton number breaking}
\label{sec:dynamical-breaking}

Let us now dive into the case of dynamical generation of this $\mu$ parameter, which allows for a richer phenomenology. 
As a well-motivated completion of the above scheme, we now turn to the dynamical version where $B-L$ is promoted to a spontaneously broken symmetry within the \SM framework.
In order to do this, we add a complex scalar singlet $\sigma$, even under $\mathcal{Z}_2$ and with $B-L=1$ defined by the last term of the relevant scalar potential, given by 
\begin{eqnarray} 
\mathcal{V}_{(s)} & = & -m^2 \Phi^\dagger \Phi \,+ \, \frac{\lambda_\Phi}{2}\left( \Phi^\dagger \Phi\right)^2
\,- \, m_\xi^2 \xi^* \xi \,+\,\frac{\lambda_\xi}{2}\left( \xi^* \xi\right)^2 
\, - \, m_\sigma^2 \sigma^* \sigma \, + \,  \frac{\lambda_\sigma}{2}\left( \sigma^* \sigma \right)^2 \nonumber \\
& + &  \lambda_{\Phi\sigma} \left(\Phi^\dagger \Phi \right)\left(\sigma^* \sigma \right)
\, + \, \lambda_{\xi\sigma} \left(\xi^* \xi \right)\left(\sigma^* \sigma \right)
\,+ \, \lambda_{\Phi\xi} \left(\Phi^\dagger \Phi \right)\left(\xi^* \xi \right) 
\,+\,\frac{\lambda_5}{2} \left(\xi^* \sigma \right)^2 \, + \, h.c. 
\label{eq:scapot}
\end{eqnarray}
The lepton number symmetry is then broken spontaneously by the vev of this complex singlet $\sigma$. The Yukawa Lagrangian in this case is the in Eq.~\eqref{eq:Yukawa-Lagrangian}. 
We assume that this breaking happens at a relatively low scale~\cite{Gonzalez-Garcia:1988okv}. %
The ``dark'' symmetry $\mathcal{Z}_2$ remains exactly conserved, enforcing the stability of the lightest ``odd'' particle.
As discussed later, the latter does meet the requirements to play the role of WIMP dark matter candidate.
In order to obtain the mass spectrum for the scalars after gauge and ${B-L}$ symmetry breaking, we expand the scalar fields as 
\begin{align}
\phi^0&=\frac{1}{\sqrt{2}}(v_\Phi + R_1+i I_1),\\
\sigma&=\frac{1}{\sqrt{2}}(v_\sigma+R_2+i I_2)\\
\xi&=\frac{1}{\sqrt{2}}(\xi_R+i \xi_I)
\end{align}
The scalar sector resulting from \eqref{eq:scapot} leads to two massive neutral CP-even scalars $h,H$ and a physical massless Goldstone boson, namely the majoron $J=\text{Im}\,\sigma$.
The mass matrix of  CP-even Higgs  scalars in the basis $(R_1,R_2)$ reads as~\cite{Joshipura:1992hp} 
\begin{align}
M_R^2=
\begin{bmatrix}
\lambda_\Phi v_\Phi^2  &  \lambda_{\Phi\sigma}v_\Phi v_\sigma \\
\lambda_{\Phi\sigma} v_\Phi v_\sigma & \lambda_\sigma v_\sigma^2 
\end{bmatrix}
\end{align}
with the mass eigenvalues given by 
\begin{align}
m_{h}^2 &=\frac{1}{2}\left(\lambda_\Phi v_\Phi^2 +\lambda_\sigma v_\sigma^2 - \sqrt{(\lambda_\sigma v_\sigma^2 - \lambda_\Phi v_\Phi^2)^2+(2\lambda_{\Phi \sigma}v_\Phi v_\sigma)^2}\right)\\
m_{H}^2 &=\frac{1}{2}\left(\lambda_\Phi v_\Phi^2 +\lambda_\sigma v_\sigma^2 + \sqrt{(\lambda_\sigma v_\sigma^2 - \lambda_\Phi v_\Phi^2)^2+(2\lambda_{\Phi \sigma}v_\Phi v_\sigma)^2}\right)
\end{align}
Under the convention $m_{h}^2\leq m_{H}^2$ the $h$ scalar must be identified with Standard Model Higgs boson.
The two mass eigenstates $h,H$ are related with the $R_1, R_2$ fields through the rotation matrix $O_R$ as,  
\begin{align}
\begin{bmatrix}
h\\
H\\
\end{bmatrix}
=O_R
\begin{bmatrix}
R_1\\
R_2\\
\end{bmatrix}
=
\begin{bmatrix}
\cos\theta & \sin\theta \\
-\sin\theta & \cos\theta \\
\end{bmatrix}
\begin{bmatrix}
R_1\\
R_2\\
\end{bmatrix},
\label{mixing relation}
\end{align}
where $\theta$ is the mixing angle in the CP even Higgs sector.
We can also write the potential parameters $\lambda_\Phi$, $\lambda_\sigma$, $\lambda_{\Phi\sigma}$ in terms of mixing angle $\theta$ and the scalar masses $m_{h,H}$ as follows
\begin{align}
\lambda_\Phi &=\frac{m_{h}^2\cos^2\theta+m_{H}^2\sin^2\theta}{v_\Phi^2}
\label{eq:lamPhi}
\end{align}
\begin{align}
\lambda_\sigma &=\frac{m_{h}^2\sin^2\theta+m_{H}^2\cos^2\theta}{v_\sigma^2}
\label{eq:lamPhi-sigma}
\end{align}
\begin{align}
\lambda_{\Phi\sigma} &=\frac{\sin 2\theta (m_{h}^2-m_{H}^2)}{2 v_\Phi v_\sigma}.
\label{eq:lam-phi-sigma}
\end{align}
From Eqs.~\eqref{eq:lamPhi}-\eqref{eq:lam-phi-sigma} one sees how the knowledge of $m_H$, $v_\sigma$ and $\sin\theta$ determines the quartic couplings
$\lambda_\Phi$, $\lambda_\sigma$ and $\lambda_{\Phi\sigma}$.
The masses of the real and imaginary components of the complex field $\xi$ are given by  
\begin{align}
m_{\xi_R}^2&=m_\xi^2+\frac{\lambda_{\Phi\xi}}{2}v_\Phi^2+\frac{\lambda_{\xi\sigma}+\lambda_5}{2}v_\sigma^2\\
  m_{\xi_I}^2&=m_\xi^2+\frac{\lambda_{\Phi\xi}}{2}v_\Phi^2+\frac{\lambda_{\xi\sigma}-\lambda_5}{2}v_\sigma^2
\label{eq:limit}
\end{align}

\begin{figure}[htb!]
\centering
\includegraphics[width=0.4\textwidth]{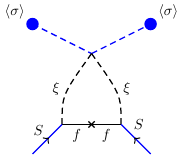}
\caption{``Scotogenic'' origin for the lepton number violating $\mu$-term in the dynamical inverse seesaw mechanism.}
\label{fig:scoto-loop}
\end{figure}

Notice that the masses in Eq.~(\ref{eq:limit}) become degenerate as $\lambda_5 v_\sigma^2 \to 0$, and this limit restores lepton number conservation. 
Indeed, in this case of dynamical breaking the parameter $\mu$ characteristic of the inverse seesaw is generated through the diagram shown in Fig.~\ref{fig:scoto-loop}, and is given as 
\begin{align}
\mu&=\frac{1}{16\pi^2}Y_\xi\mathcal{M}_f\left( \frac{m_{\xi_R}^2}{m_{\xi_R}^2-\mathcal{M}_{f}^2}\log \left( \frac{m_{\xi_R}^2}{\mathcal{M}_{f}^2}\right) \,-\, \frac{m_{\xi_I}^2}{m_{\xi_I}^2-\mathcal{M}_{f}^2}\log \left( \frac{m_{\xi_I}^2}{\mathcal{M}_{f}^2} \right) \right)Y_\xi^T
\end{align}
Assuming for simplicity that $Y_\xi$ and $\mathcal{M}_f$ are diagonal, we can factorize the above loop functions as follows: 
\begin{equation}
\mu = \left( \begin{array}{ccc}
\mu_1 & 0 & 0 \\
0 & \mu_2 & 0 \\
0 & 0 & \mu_3 
\end{array} \right) \, , \quad \mu_i=\frac{Y_{\xi}^{(i)2}M_f^{(i)}}{16\pi^2}\left( \frac{m_{\xi_R}^2}{m_{\xi_R}^2-M_f^{(i)2}}\log \left( \frac{m_{\xi_R}^2}{M_f^{(i)2}}\right) \,-\, \frac{m_{\xi_I}^2}{m_{\xi_I}^2-M_f^{(i)2}}\log \left( \frac{m_{\xi_I}^2}{M_f^{(i)2}} \right) \right),
\end{equation}
Note that $m_{\xi_R}^2=m_{\xi_I}^2$ leads to an exact cancellation between the $\xi_R$ and $\xi_I$ loops, and vanishing neutrino masses,
just as in the simplest scotogenic model.
Moreover, under the approximation $M_f^{(i)2},m_{\xi_R}^2$ and $M_f^{(i)2}-m_{\xi_R}^2\gg \lambda_5 v_\sigma^2$, the parameter $\mu$ can be given as follows,
\begin{align}
\mu_i\approx \frac{1}{16\pi^2}\frac{\lambda_5 v_\sigma^2}{M_f^{(i)2}-m_{\xi_R}^2}M_f^{(i)}Y_{\xi}^{(i)2}
\end{align}
Once $\mu$ is generated this induces the following small neutrino masses:
\begin{eqnarray}
\mathcal{M}_\nu &= m_D M^{-1}\mu M^{-1T}m_D^T\equiv m_D \mathcal{M}_R^{-1} m_D^{T}
\label{eq:neutrino-mass}
\end{eqnarray}
where we have defined $\mathcal{M}_R^{-1}=M^{-1}\mu M^{-1T}$. From the above equations it is clear that the smallness of neutrino mass will be symmetry- as well as loop-protected. 
We choose to work in a basis where $M$ is diagonal, so that $\mathcal{M}_R$ is also diagonal.
As in~\cite{Deppisch:2004fa} we parameterize the Yukawa coupling using the Casas-Ibarra form~\cite{Casas:2001sr}, 
\begin{align}
Y_{\nu^c}=\frac{\sqrt{2}}{v_\Phi}U^{\dagger}_{\text{lep}}\sqrt{\widehat{\mathcal{M}}_\nu} R \sqrt{\mathcal{M}_R}
\label{eq:Ynu}
\end{align}
Here $R$ is a $3 \times 3$ complex matrix such that $R R^T = \mathbb{I}_{3}$, where $\mathbb{I}_{3}$ is the $3 \times 3$ unit matrix, and the neutrino mass matrix is diagonalized as 

\begin{equation}
U_{\text{lep}}^{T} \, \mathcal{M}_\nu \, U_{\text{lep}}=\widehat{\mathcal{M}}_\nu\equiv
\left(
\begin{array}{ccc}
m_1 & 0 & 0\\
0 & m_2 & 0\\
0 & 0 & m_3
\end{array}
\right) \, ,\quad \text{\bf for NO}
\label{eq:mnudiagNO}
\end{equation}
\begin{equation}
U_{\text{lep}}^{T} \, \mathcal{M}_\nu \, U_{\text{lep}}=\widehat{\mathcal{M}}_\nu\equiv
\left(
\begin{array}{ccc}
m_3 & 0 & 0\\
0 & m_2 & 0\\
0 & 0 & m_1
\end{array}
\right) \, ,\quad \text{\bf for IO}
\label{eq:mnudiagIO}
\end{equation}
where $U_{\text{lep}}$ is the leptonic mixing matrix and $m_i$'s are the light neutrino masses. The form of the complex matrix $R$ is given as follows,
\begin{align}
R = \left( \begin{array}{ccc}  1 & 0 & 0
  \\ 0 & \cos x & \sin x \\
0 & -\sin x & \cos x\\
\end{array} \right)  \left( \begin{array}{ccc}  \cos y & 0 & \sin y
  \\ 0 & 1 & 0 \\
-\sin y & 0 & \cos y\\
\end{array} \right)  \left( \begin{array}{ccc}  \cos z & \sin z & 0
  \\ -\sin z & \cos z & 0 \\
0 & 0 & 1\\
\end{array} \right)
\end{align}
where $x,y$ and $z$ are the complex angles. 
\section{Higgs physics at colliders}
\label{sec:Higgs-physics}

In this section we discuss the constraints on the relevant parameter space of Higgs bosons which follow from collider searches performed at the LHC.
First, notice that, due to the presence of this heavy Higgs $H$, the coupling of the standard Higgs boson to \sm particles
gets modified according to the substitution rule 
\begin{align}
h_{\text{SM}}\to \cos\theta \,h - \sin\theta\, H
\label{eq:substitution}
\end{align}
Moreover, due to the spontaneous breaking of $U(1)_{B-L}$, there exist a massless Nambu-Goldstone boson, called majoron $J$. 
Since the scale associated with lepton number violation is relatively low, the \sm Higgs boson $h$ can have potentially large invisible decays~\cite{Joshipura:1992hp}:
\begin{align}
\Gamma(h\to JJ)&=\frac{1}{32\pi m_h}\frac{m_h^4\sin^2\theta}{v_\sigma^2}\sqrt{1-\frac{4m_{J}^2}{m_h^2}}
\label{eq:inv-majoron}
\end{align}

In addition to the majoron channel, if either of $m_{\xi_R}$ or $m_{\xi_I}$ is smaller than half of the Higgs mass, then these two channels will also contribute to the invisible decay. 
The partial decay width to $\xi_{R}\xi_R$ and $\xi_I\xi_I$ are given as
\begin{align}
\Gamma(h\to\xi_R\xi_R)&=\frac{1}{32\pi m_h}\big((\lambda_{\xi\sigma}+\lambda_5)v_\sigma\sin\theta + \lambda_{\Phi\xi}v_\Phi\cos\theta \big)^2\sqrt{1-\frac{4m_{\xi_R}^2}{m_h^2}}\\
\Gamma(h\to\xi_I\xi_I)&=\frac{1}{32\pi m_h}\big((\lambda_{\xi\sigma}-\lambda_5)v_\sigma\sin\theta + \lambda_{\Phi\xi}v_\Phi\cos\theta \big)^2\sqrt{1-\frac{4m_{\xi_I}^2}{m_h^2}}
\label{eq:inv-dark}
\end{align}
Hence the total invisible decay width of \sm Higgs boson $h$ is given as 
\begin{align}
\Gamma^{\text{inv}}(h)=\Gamma(h\to JJ)+\Gamma(h\to\xi_R\xi_R)+\Gamma(h\to\xi_I\xi_I)
\label{eq:inv-Higgs}
\end{align}
Accordingly, the invisible branching ratio for $h$ is given by
\begin{align}
\text{BR}^{\text{inv}}(h)=
\frac{\Gamma^{\text{inv}}(h)}{\cos^2\theta\Gamma^\text{SM}(h)+\Gamma^{\text{inv}}(h)}
\label{eq:inv-BR_Higgs}
\end{align}

The collider implications of invisibly decaying Higgs bosons have been extensively discussed in theory papers~\cite{Romao:1992zx,Eboli:1994bm,DeCampos:1994fi,Romao:1992dc,deCampos:1995ten,
  deCampos:1996bg,Diaz:1998zg,Hirsch:2004rw,Hirsch:2005wd,Bonilla:2015uwa,Bonilla:2015jdf}.
Recent experimental studies have been given by the LHC collaborations~\cite{CMS:2018yfx,ATLAS:2019cid}. 
Invisible Higgs decays will also be an interesting point in the agenda of planned lepton collider experiments such as CEPC, FCC-ee, ILC and CLIC~\cite{CEPCStudyGroup:2018ghi,FCC:2018evy,Bambade:2019fyw,deBlas:2018mhx}. \\[-.4cm] 

Due to the presence of the invisible Higgs decays~(Eq.~\eqref{eq:inv-Higgs}) and modified Higgs coupling~(Eq.\eqref{eq:substitution}) the branching ratio to \sm final state particles gets modified as
\begin{align}
\text{BR}_f(h)=
\frac{\cos^2\theta \Gamma_f^\text{SM}(h)}{\cos^2\theta \Gamma^\text{SM}(h)+\Gamma^{\text{inv}}(h)}
\label{eq:BR-visible}
\end{align}

Since the coupling of $h$ and $H$ to other \sm fermions and gauge bosons are suppressed relative to the standard values by $\cos\theta$ and $\sin\theta$, the
$h$ or $H$ production cross-sections are also modified as,  
\begin{align}
\sigma(pp\to h) &= \cos^2\theta\sigma^{\text{SM}}(pp\to h) \\
\sigma(pp\to H) &= \sin^2\theta\sigma^{\text{SM}}(pp\to H ) 
\end{align}
where $\sigma^{\text{SM}}(pp\to h)$ and $\sigma^{\text{SM}}(pp\to H)$ are the \sm cross-sections for Higgs production for $m_{h}$ and $m_{H}$. 
Notice that the factors $\cos^2\theta$ or $\sin^2\theta$ with respect to the conventional ones.\\[-.4cm] 

Now we are ready to discuss the constraints on the relevant parameter space of Higgs bosons which follow from searches performed at LHC.
First of all, these come from the current upper limit on the branching ratio to invisible decay modes by the CMS experiment~\cite{CMS:2018yfx}
~\footnote{The present bound on invisible Higgs decays from ATLAS is $\text{BR}(h\to \text{Inv})\leq 0.26$~\cite{ATLAS:2019cid}.}, 
\begin{align}
\text{BR}^{\text{inv}}(h) \leq 0.19.
\label{eq:invisible}
\end{align}
\begin{figure}[htb!]
\centering
\includegraphics[width=0.5\textwidth]{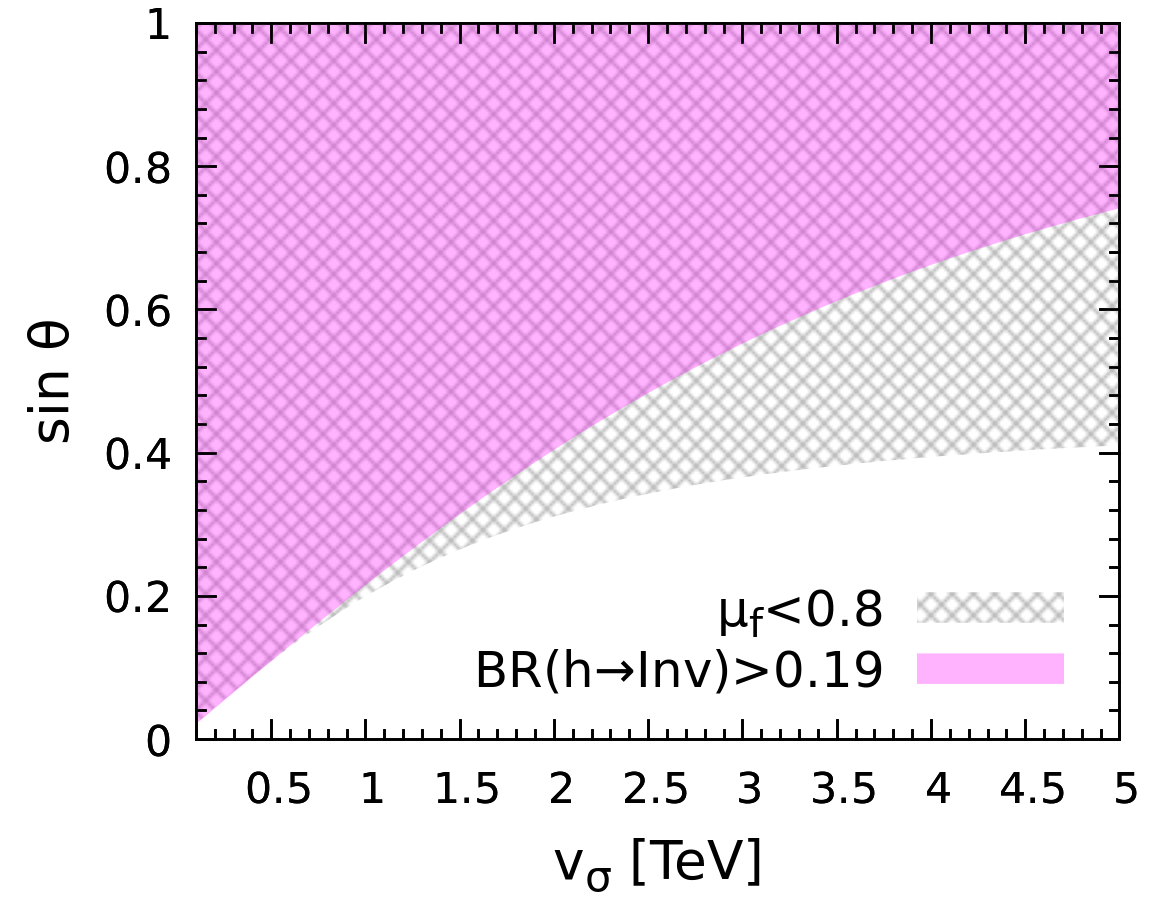}
\caption{The shaded areas on $\sin\theta$ versus $v_\sigma$ are ruled out by the present limit on the invisible Higgs decay from Eq.~(\ref{eq:invisible}) (magenta) and
  also from the constraints on the signal strength parameter $\mu_f$ in Eq.~(\ref{eq:muf})~(gray), under the assumption that $m_{\xi_{R,I}}>m_h/2$.}
\label{fig:limit-from-inv-mu}
\end{figure}
If the dark matter mass $m_{\xi}>m_h/2$, then only the majoron channel $\Gamma(h\to JJ)$ is open. 
The invisible Higgs decay constraint can be translated as an upper bound on the $\sin\theta-v_\sigma$ plane, shown in the Fig.~\ref{fig:limit-from-inv-mu},
where the magenta region is excluded from the current invisible Higgs decay limit, Eq.~\ref{eq:invisible}. \\[-.4cm] 

Notice that in the opposite case of $m_{\xi}<m_h/2$, the decay modes $\Gamma(h\to\xi_{R/I}\xi_{R/I})$ will also contribute to the invisible Higgs decay, Eq.~(\ref{eq:inv-Higgs}). 
These depend on other quartic couplings such as $\lambda_{\Phi\xi}$, $\lambda_{\xi\sigma}$ and $\lambda_5$. 
In this case, neglecting $\sin\theta\sim 0$, the invisible Higgs decay constraint can be translated as an upper bound on the quartic coupling $\lambda_{\Phi\xi}$,
(see also magenta region in Fig.~\ref{fig:fitdm} below): 
\begin{align}
\lambda_{\Phi\xi}\left(1-\frac{4m_\xi^2}{m_h^2}\right)^{\frac{1}{4}}\leq 9.8\times 10^{-3}.
\end{align}

Apart from the above invisible Higgs decay constraints, we also have the LHC measurements of several visible decay modes of the \sm Higgs boson, $h$. 
These are given in terms of the so-called signal strength parameters, 
\begin{align}
\mu_f &=\frac{\sigma^{\text{NP}}(pp\to h)}{\sigma^{\text{SM}}(pp\to h)} \frac{\text{BR}^{\text{NP}}(h\to f)}{\text{BR}^{\text{SM}}(h\to f)} \nonumber  \\
&=\frac{\sigma^{\text{NP}}(pp\to h)}{\sigma^{\text{SM}}(pp\to h)} \frac{\Gamma^{\text{NP}}(h\to f)}{\Gamma^{\text{SM}}(h\to f)} \frac{\Gamma^{\text{SM}}(h\to \text{all})}{\Gamma^{\text{NP}}(h\to \text{all})}
\end{align}
where $\sigma$ is the cross-section for Higgs production, NP and SM stand for new physics and \sm respectively. 
For the 8TeV data, there are combined results for signal strength parameters from combined ATLAS and CMS analyses~\cite{TheATLASandCMSCollaborations:2015bln}, which we list in Table.~\ref{tab:1}. 
\begin{table}[!h]
\centering
\begin{tabular}{|c | c | c | c |}
\hline
\hspace{0.25cm} Channel \hspace{0.25cm} &  \hspace{0.25cm} ATLAS  \hspace{0.25cm} & \hspace{0.35cm} CMS  \hspace{0.35cm} & \hspace{0.25cm}ATLAS+CMS \hspace{0.25cm} \\
\hline
$\mu_{\gamma\gamma}$   &
$1.15^{+0.27}_{-0.25}$   &
$1.12 ^{+0.25}_{-0.23}$  &
$1.16^{+0.20}_{-0.18}$
\\*[2mm]
$\mu_{WW}$   &
$1.23^{+0.23}_{-0.21}$  &
$0.91^{+0.24}_{-0.21}$   &
$1.11^{+0.18}_{-0.17}$
\\*[2mm]
$\mu_{ZZ}$   &
$1.51^{+0.39}_{-0.34}$ &
$1.05^{+0.32}_{-0.27}$  &
$1.31^{+0.27}_{-0.24}$
\\*[2mm]
$\mu_{\tau\tau}$  &
$1.41^{+0.40}_{-0.35}$ &
$0.89^{+0.31}_{-0.28}$ &
$1.12^{+0.25}_{-0.23}$
\\*[2mm]
\hline
\end{tabular}
\caption{\label{tab:1} Combined ATLAS and CMS results for the 8 TeV data, Ref.~\cite{TheATLASandCMSCollaborations:2015bln}.} 
\end{table}

For the 13TeV Run-2, there is no combined final data so far, and the available data is separated by production processes. 
Table~\ref{tab:2} compiles the recent results from ATLAS~\cite{ATLAS:2019nkf}.
\begin{table}[ht]
\begin{center}
\begin{tabular}{|c|c|c|c|c|}
    \hline
 \hspace{0.25cm} Decay \hspace{0.25cm} &\multicolumn{4}{c|}{Production Processes}\\[+2mm]\cline{2-5}
  Mode & \hspace{0.35cm} \texttt{ggF} \hspace{0.35cm} & \hspace{0.35cm} \texttt{VBF} \hspace{0.35cm} & \hspace{0.35cm} \texttt{VH} \hspace{0.35cm} & \hspace{0.35cm} \texttt{ttH} \hspace{0.35cm}  \\[+2mm]
  \hline
  \vb{18}   $h\to \gamma\gamma$
         &$0.96^{+0.14}_{-0.14}$
         &$1.39^{+0.40}_{-0.35}$ 
         &$1.09^{+0.58}_{-0.54}$
         &$1.10^{+0.41}_{-0.35}$ \\[+2mm]
   \vb{18}   $h\to ZZ$
         &$1.04^{+0.16}_{-0.15}$
         &$2.68^{+0.98}_{-0.83}$
         &$0.68^{+1.20}_{-0.78}$
         &$1.50^{+0.59}_{-0.57}$ \\[+2mm]
  \vb{18}  $h\to WW$
         &$1.08^{+0.19}_{-0.19}$
         &$0.59^{+0.36}_{-0.35}$
         &$-$
         &$1.50^{+0.59}_{-0.57}$ \\[+2mm] 
\hline
 \vb{18}   $h\to \tau\tau$
         &$0.96^{+0.59}_{-0.52}$
         &$1.16^{+0.58}_{-0.53}$
         &$-$
         &$1.38^{+1.13}_{-0.96}$ \\[+2mm]
   \vb{18}   $h\to bb$
          &$-$
          &$3.01^{+1.67}_{-1.61}$ 
          &$1.19^{+0.27}_{-0.25}$
          &$0.79^{+0.60}_{-0.59}$ \\[+2mm]
  \hline
\end{tabular}
\end{center}
\caption{ATLAS results for 13 TeV data, taken from Ref.~\cite{ATLAS:2019nkf}}
\label{tab:2}
\end{table}
One sees that, although compatible at $1\sigma$, current limits still have quite large errors. For simplicity, we adopt the conservative range,
\begin{align}
0.8\leq \mu_f\leq 1.2.
\label{eq:muf}
\end{align}
The gray shaded region in Fig.~\ref{fig:limit-from-inv-mu} is excluded from the constraint in Eq.~\eqref{eq:muf} under the assumption that only $h\to JJ$ invisible mode is open.
As can be seen from Fig.~\ref{fig:limit-from-inv-mu}, for low values of $v_\sigma$ up to around 1 TeV, both~Eq.~\eqref{eq:invisible} and Eq.~\eqref{eq:muf} lead to similar limits on $\sin\theta$.  
However, for $v_\sigma > 1$ TeV the limit from Eq.~\eqref{eq:invisible} gets relaxed since, the larger the $v_\sigma$, the smaller the invisible decay mode $h\to JJ$.
As a result, for larger $v_\sigma$ values the Higgs invisible decay gives a weaker exclusion limit on $\sin\theta$ than that coming from $\mu_f$.
Again we stress that for $m_{\xi}\leq m_h/2$, the exclusion region will depend on the values of other quartic couplings such as $\lambda_{\Phi\xi}$, $\lambda_{\xi\sigma}$.\\[-.4cm] 

Additional constraints will come from the direct search of the heavy CP-even Higgs boson $H$. 
Two types of LHC searches are relevant, namely, direct Higgs production $pp\to H$ with successive decay to \sm particles
e.g., $WW$, $ZZ$ and subsequent decays $WW\to 2\ell 2\nu$ and $ZZ\to 4\ell$)~\cite{CMS:2017dua, ATLAS:2017uhp}.
Alternatively, it can also decay to a pair of \sm Higgs bosons $h$ if kinematically allowed.
These has been discussed, for example, in Ref.~\cite{Mandal:2021acg}. 
We find that with $v_\sigma\gg v_{H}$ the constraint from the signal strength parameter is stronger. 
\section{Phenomenology of dark matter} 
\label{sec:dark-matter}

In this section we present the results of our analysis of dark matter phenomenology.
In addition to ensuring radiative generation of neutrino masses, the $\mathcal{Z}_2$ symmetry in the dark sector is responsible for the stability of ``lightest dark particle'' (LDP). 
There are three LDP options, dark fermion $f$ and the real or imaginary parts of the $\xi$ scalar, $\xi_R$ or $\xi_I$. 
In our analysis, we assume scalar dark matter $\xi_R$, with the condition $\lambda_5< 0$~(the opposite scenario with $\lambda_5>0$ would have $\xi_I$ as the dark matter particle)
~\footnote{The fermionic dark matter case is more restrictive, as the parameters entering in the evaluation of the relic density are directly related to those determining the neutrino masses,
  in contrast to the scalar dark matter case, where direct detection involves the Higgs portal.}.\\[-.4cm]  

In order to calculate all the vertices, mass matrices, tadpole equations etc the model is implemented in the SARAH package~\cite{Staub:2015kfa}.
On the other hand, the thermal component of the dark matter relic abundance is determined using micrOMEGAS-5.0.8~\cite{Belanger:2018mqt}.
\subsection{Relic density}  
\label{sec:relic-density}
There are several dark matter annihilation and coannihilation diagrams which will contribute to the relic abundance of our assumed dark matter, $\xi_R$. 
The relic density of dark matter $\xi_R$ is mostly determined by CP-even scalars-mediated s-channel annihiliation to \sm
final states~($\ell^+\ell^-$, $q\bar{q}$, $W^+W^-$, $ZZ$, $\gamma\gamma$, $hh$) as well as to $HH$ and $JJ$ final states. 
A sub-dominant role is played by annihiliation into $hh,HH$ and $JJ$ via the direct 4-scalar vertices $h^2\xi_R^2$, $H^2\xi_R^2$ and $J^2\xi_R^2$, respectively.
Also there could be additional contributions from $\xi_{R/I}$ exchange in the t-channel.  
In Fig.~\ref{fig:annihilation-diagram} of Appendix~\ref{app:annihilation} we collect the Feynman diagrams contributing to $\xi_R$ annihilations and co-annihilations. 
In Table.~\ref{tab:cubic-coupling} and \ref{tab:quartic-coupling} of Appendix~\ref{app:annihilation},
we have listed the cubic and quartic scalar interactions which play a role in the annihilation channels. 
Using these one can see that dark matter annihilation is mainly determined by mixed quartic couplings, such as $\lambda_{\Phi\xi}$, $\lambda_{\xi\sigma}$, $\lambda_5$ and $\lambda_{\Phi\sigma}$. 
The $\lambda_5$ dependence is negligible, as $\lambda_5$ must be very small in order to generate light neutrino masses with sizeable charged lepton flavour violation (see Sec.~\ref{sec:LFV}). 
One also sees from Eq.~\ref{eq:lam-phi-sigma} that the mixed quartic coupling $\lambda_{\Phi\sigma}$ is equivalent to the mixing angle $\sin\theta$ once we fix the free parameters $m_H$ and $v_\sigma$.
Hence non-zero $\sin\theta$ implies non-zero $\lambda_{\Phi\sigma}$. 
As we discussed in sec.~\ref{sec:Higgs-physics} the mixing angle $\sin\theta$ is tightly constrained from the LHC experiments. 
In order to illustrate the main features of our chosen dark matter candidate $\xi_R$ we choose the following benchmark points:
\begin{align}
&\text{\textbf{BP1:} }\sin\theta =0,  \lambda_{\Phi\xi}=0.01,\,\,\,
\text{\textbf{BP2:} } \sin\theta =0, \lambda_{\Phi\xi}=0.1. \\
&\text{\textbf{BP3:} } \sin\theta =0.1,  \lambda_{\Phi\xi}=0.01,\,\,\,
\text{\textbf{BP4:} }\sin\theta =0.1, \lambda_{\Phi\xi}=0.1.
\end{align}
while fixing other parameters as $m_H=1\,\text{TeV}, v_\sigma=3 \text{ TeV}, \lambda_\xi=0.1$ and $\lambda_{\xi\sigma}=0.1$. 
%
\begin{figure}[htb!]
\includegraphics[width=0.55\textwidth]{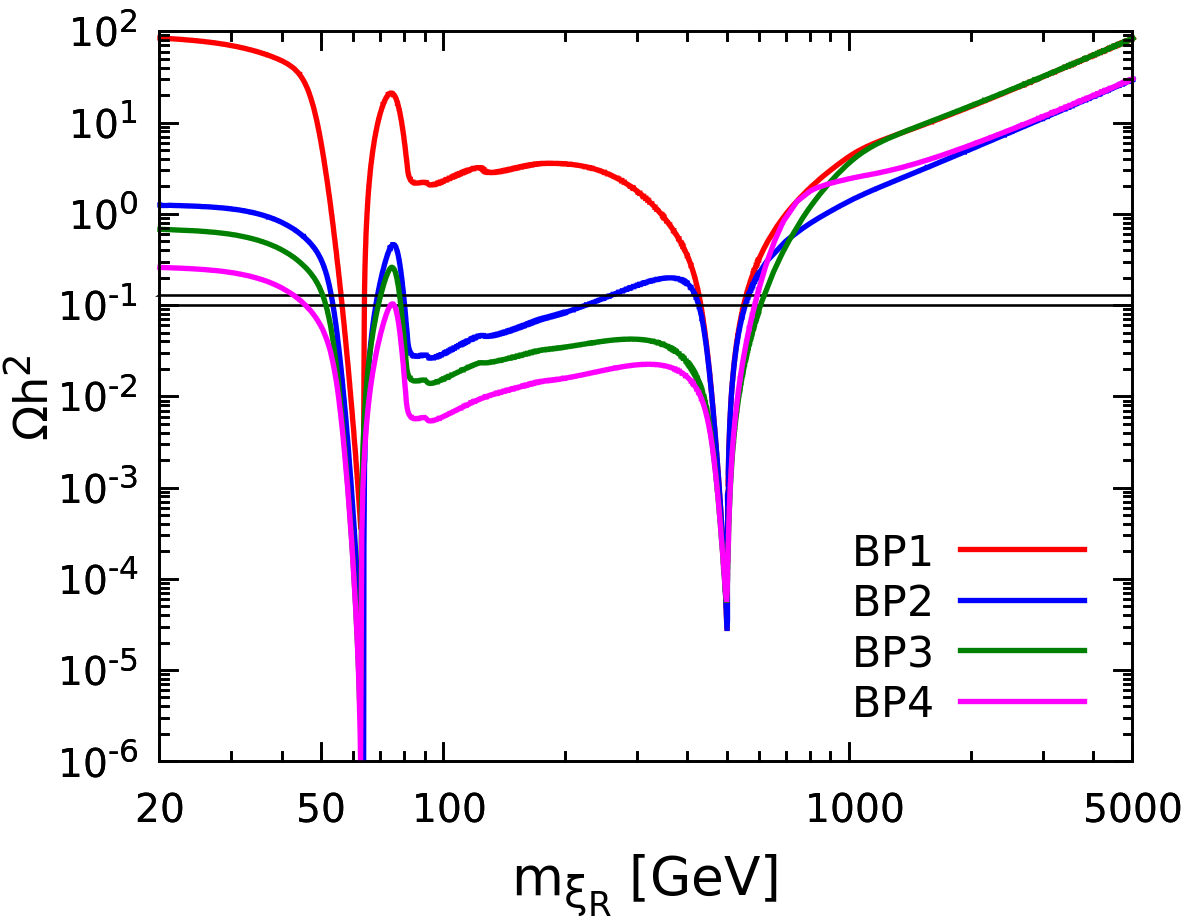}
\caption{ Relic abundance as a function of the dark matter mass $m_{\xi_R}$.
  The narrow horizontal black lines delimit the measured $3\sigma$ range given by Planck satellite data, Eq.~\eqref{eq:Pl}.}
\label{fig:relic}
\end{figure}

The relic density of the dark matter candidate $\xi_R$ as a function of its mass is shown in Fig.~\ref{fig:relic}. 
The calculation is performed for the four benchmarks \textbf{BP1}~(red line), \textbf{BP2}~(blue line), \textbf{BP3}~(green line) and \textbf{BP4}~(magenta line).
The narrow horizontal band is the $3\sigma$ range for cold dark matter derived from the Planck satellite data~\cite{Planck:2018vyg}:
\begin{align}
0.1126 \leq \Omega_{\xi_R} h^2 \leq 0.1246.
  \label{eq:Pl}
\end{align}
Only for solutions falling exactly within this band the totality of the dark matter can be explained by $\xi_R$.  

Various features of the relic density in Fig.~\ref{fig:relic} can be understood by looking in detail into the $\xi_R$ annihilation channels
shown in fig.~\ref{fig:annihilation-diagram} of Appendix~\ref{app:annihilation}. 
The two dips at $m_{\xi_R}\sim m_h/2$ and $m_{\xi_R}\sim m_H/2$ correspond to annihilation via s-channel $h$ and $H$ exchange. 
These becomes very efficient when the SM-like Higgs boson $h$ or heavy Higgs $H$ are on-shell, precluding us from obtaining a relic density matching Planck observations. 
For $m_{\xi_R} \gsim 80$ GeV, annihilations of $\xi_R$ into $W^+W^-$ and $ZZ$ are particularly important, thus explaining the drop in the relic abundance for $m_{\xi_R}\sim 80$ GeV.
For very heavy $m_{\xi_R}$ the relic density increases due to the suppressed annihilation cross section, which drops as $\sim\frac{1}{m_{\xi_R}^2}$.  
The annhiliation cross section also increases with increasing $\lambda_{\Phi\xi}$ or $\sin\theta$.  This way we understand relative relic densities for different benchmarks. 
Notice also that for large $m_{\xi_R}$, the relic densities for fixed $\lambda_{\Phi\xi}$ coincide. 
This happens as for large $m_{\xi_R}$, the Higgs mediated annihiliation channel $\xi_R\xi_R\to W^+W^-, ZZ$ dominates, and is determined by the mixed quartic coupling $\lambda_{\Phi\xi}$
if $\lambda_{\xi\sigma}$ is fixed and $\sin\theta$ is small. 
Note that, due to the existence of the majoron $J$, one can have significant annihilation to the final state $JJ$, depending on the mixed quartic couplings strength.
Notice also that, when the magnitude of the mass difference $m_{\xi_R}^2-m_{\xi_I}^2$ is small, one must take into account the contributions from both species $\xi_R,\xi_I$.
This can have a significant impact in lowering the relic dark matter density. 

\subsection{Direct detection}
\label{sec:direct-detection}
\begin{figure}[htb!]
\centering
\includegraphics[width=0.3\textwidth]{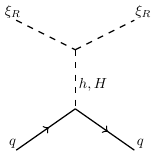}
\caption{Higgs-mediated tree-level Feynman diagrams contributing to the scattering of $\xi_R$ off nuclei.}
\label{fig:dd-higgs}
\end{figure}

Let us now study the direct detection prospects of our scalar dark matter candidate $\xi_R$. 
In our model, the elastic scattering of the dark matter scalar $\xi_R$ with a nucleon happens via two t-channel diagrams mediated by $h$ and $H$, see Fig.~\ref{fig:dd-higgs}.
The effective Lagrangian for nucleon-dark matter interaction can be written as 
\begin{align}
\mathcal{L}_{\text{eff}}=a_N\bar{N}N\xi_R^2
\end{align}
where $a_N$ is the effective coupling between the dark matter and the nucleon. The resulting spin-independent scattering cross section is given by~\cite{Basak:2021tnj},
\begin{align}
\sigma^{\text{SI}}=\frac{\mu_N^2 m_N^2 f_N^2}{4\pi m_{\xi_R}^2 v_\Phi^2}\Big(\frac{\lambda_{h\xi_R\xi_R}}{m_h^2}\cos\theta - \frac{\lambda_{H\xi_R\xi_R}}{m_H^2}\sin\theta\Big)^2,
\label{eq:SI}
\end{align}
where $\mu_N=\frac{m_N m_{\xi_R}}{m_N+m_{\xi_R}}$ is the reduced mass for nucleon-dark matter system. Here $f_N$ is the form factor, which depends on hadronic matrix elements.
The trilinear couplings $\lambda_{h\xi_R\xi_R}$ and $\lambda_{H\xi_R\xi_R}$ are given as
\begin{align}
\lambda_{h\xi_R\xi_R}&=\lambda_{\Phi\xi}v_\Phi\cos\theta+(\lambda_{\xi\sigma}+\lambda_5)v_\sigma\sin\theta \\
\lambda_{H\xi_R\xi_R}&=-\lambda_{\Phi\xi}v_\Phi\sin\theta+(\lambda_{\xi\sigma}+\lambda_5)v_\sigma\cos\theta 
\end{align}
where, as we have seen, $\lambda_5$ is really tiny and can be neglected.
Note that Eq.~\eqref{eq:SI} generalizes the expression corresponding to the singlet scalar dark matter in~\cite{Cline:2013gha}.
The relative sign between the $h$ and $H$ contributions arises as the coupling of the Higgs boson to \sm particles gets modified according to the substitution rule in Eq.~\eqref{eq:substitution}.
Due to the presence of two channels, depending on our parameter choice, there can be destructive interference between these channels so that direct detection can be very small. 
\begin{figure}[htb!]
\includegraphics[width=0.55\textwidth,height=0.45\textwidth]{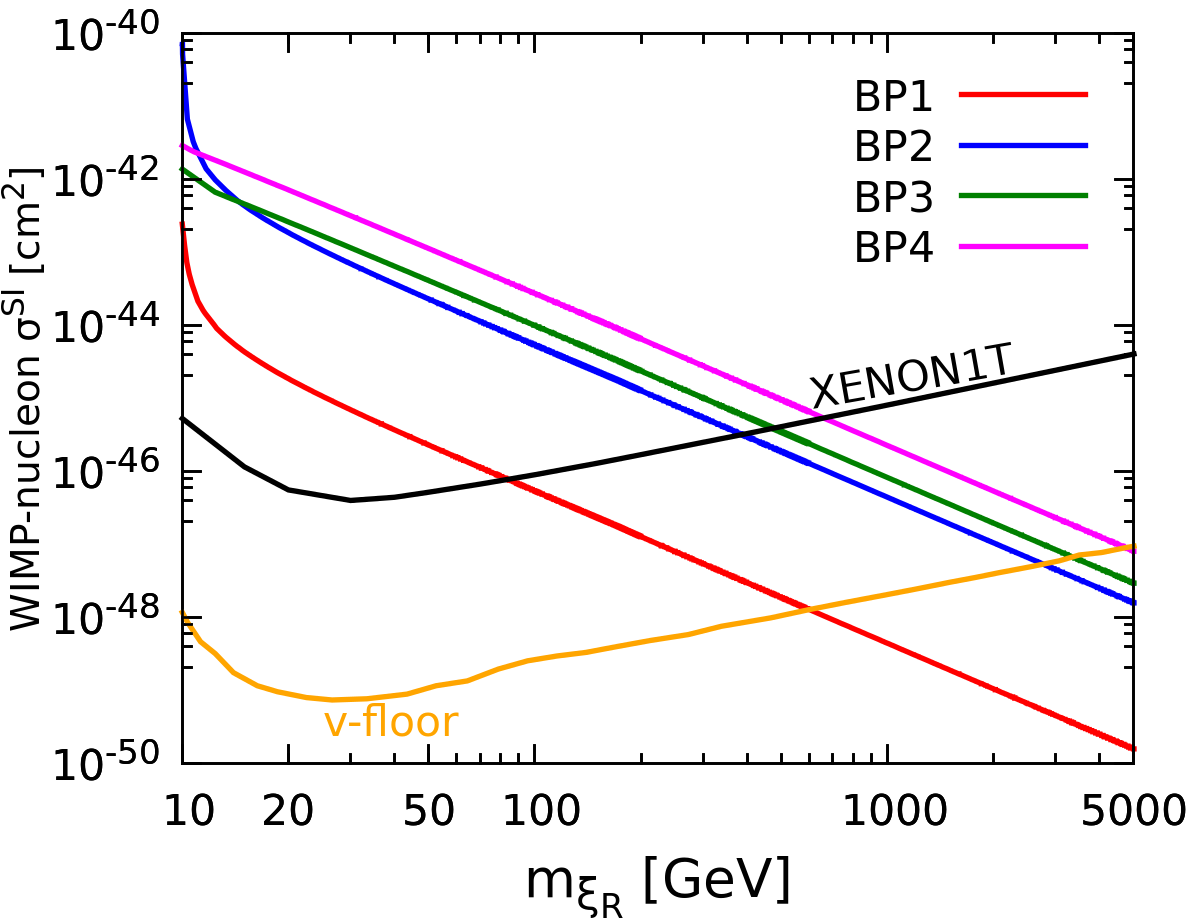}
\caption{ 
  Spin-independent $\xi_R$-nucleon cross-section versus the $\xi_R$ mass for different benchmarks.
  The solid black line denotes the recent upper bound from the XENON1T experiment~\cite{Aprile:2018dbl} while
  the orange line corresponds to the ``neutrino floor''~\cite{Billard:2013qya}.}
\label{fig:direct-detection}
\end{figure}

In Fig.~\ref{fig:direct-detection} we show the spin-independent $\xi_R$-nucleon cross-section versus the $\xi_R$ mass, for the same benchmarks as in Fig.~\ref{fig:relic}.
The black solid line denotes the latest upper bound from the XENON1T collaboration. 
Other experiments as well, such as LUX~\cite{LUX:2016ggv} and PandaX-II~\cite{PandaX-II:2016vec} give weaker (undisplayed) constraints. 
We also indicate the ``neutrino floor'' constraint from coherent elastic neutrino scattering coming from several astrophysical sources~\cite{Billard:2013qya}. 
Clearly there are high mass solutions with the correct dark matter relic density while,
for our chosen benchmarks, most of the low mass dark matter region is ruled out by the XENON1T direct detection cross section upper limits.
As we will discuss next, there are also tight constraints on low-mass dark matter from collider experiments. \\[-.4cm]

\begin{figure}[htb!]
\centering
\includegraphics[width=0.6\textwidth]{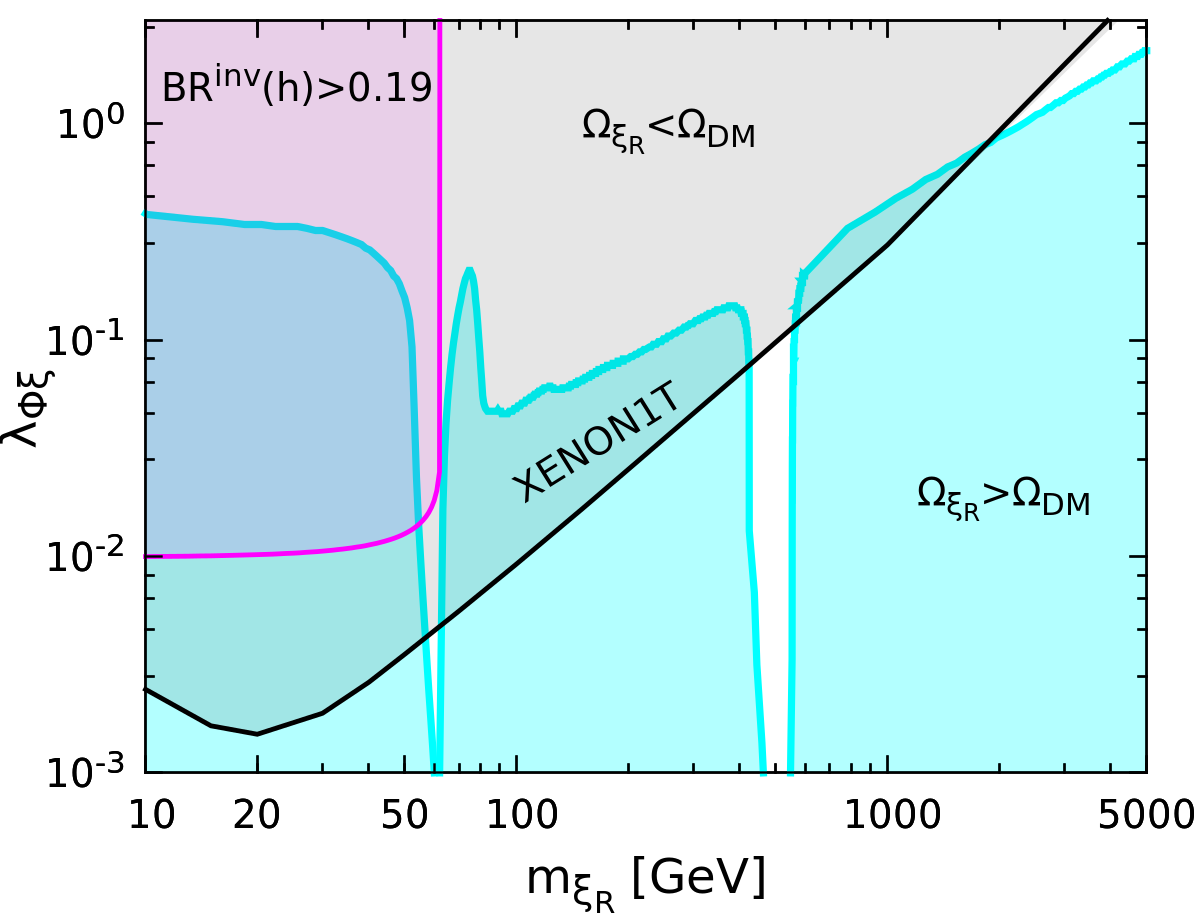}
\caption{ Allowed values of the quartic coupling $\lambda_{\Phi\xi}$ as a function of dark matter mass $m_{\xi_R}$ from relic density, direct detection and invisible Higgs decay.
  Here we take \textbf{BP1} i.e. $m_H=1\,\text{TeV}, \sin\theta=0, v_\sigma=3 \text{ TeV}, \lambda_\xi=0.1$, $\lambda_{\xi\sigma}=0.1$.
  Along the cyan line we have the correct dark matter relic abundance. The region below this line corresponds to overabundance and is ruled out.
  The region above the black line is excluded by the XENON1T direct detection limit~\cite{Aprile:2018dbl}.
  The magenta region is excluded by the invisible Higgs decay constraint from the CMS experiment~\cite{CMS:2018yfx}.}
\label{fig:fitdm}
\end{figure}

In Fig.~\ref{fig:fitdm} we put together the constraints from relic density, direct detection and invisible Higgs decay.
To illustrate how the latter plays a role in restricting the dark matter parameter space
we choose one of the previous benchmarks, \textbf{BP1}, keeping now the quartic coupling $\lambda_{\Phi\xi}$ as free parameter.  
Along the cyan line the real scalar singlet ${\xi_R}$ gives the dark matter relic abundance within $3\sigma$, Eq.~\eqref{eq:Pl}. 
The region below this cyan line gives over-abundant dark matter. The shaded black region is excluded by the XENON1T experiment~\cite{Aprile:2018dbl}. 
The shaded magenta region is excluded from invisible Higgs decay constraint, Eq.~\eqref{eq:invisible}. 
We see that this benchmark allows for our dark matter candidate with mass $m_{\xi_R}>1.8$~TeV, consistent with all existing constraints. 
However, lower dark matter masses are possible, e.g. close to the dips associated to the SM-like Higgs and the heavier CP-even scalar boson.

\section{Charged lepton flavour violation and collider searches}
\label{sec:LFV}

In the inverse seesaw mechanism charged lepton flavour violation (cLFV) is mediated by singlet neutrino exchange~\cite{Bernabeu:1987gr},
and this is also a feature of our proposed scotogenic extension.
Results for charged lepton flavour violation obtained in previous papers basically hold here~\cite{Gonzalez-Garcia:1991brm,Ilakovac:1994kj,Deppisch:2004fa,Deppisch:2005zm,DeRomeri:2016gum}.
In contrast to the original scotogenic model and most of its extensions~\cite{Ma:2006km,Hirsch:2013ola,Merle:2016scw,Rocha-Moran:2016enp,Restrepo:2019ilz,Avila:2019hhv},
the dark sector does not play a direct role in inducing cLFV. 

\subsection{The minimal \lfv hypothesis}
\label{sec:minim-lfv-hypoth}

For simplicity here we assume that the Yukawa interaction $Y_{\nu^c} \overline{L} i\tau_2 H^* \nu^c$  provides the only source of lepton flavour violation.
This implies that these parameters will be responsible both for neutrino oscillations, as well as cLFV decays such as $\ell_i\to\ell_j\gamma$, $\ell_i\to\ell_j \ell_k^+\ell_k^-$
and $\mu\to e$ conversions in nuclei.
As in the conventional inverse seesaw model, these are induced by isosinglet neutrino exchange~\cite{Bernabeu:1987gr}.
There have been many numerical studies~\cite{Gonzalez-Garcia:1991brm,Ilakovac:1994kj,Deppisch:2004fa,Deppisch:2005zm,DeRomeri:2016gum}.  
The rate for the radiative $\ell_i\to\ell_j\gamma$ decay is given by~\cite{Minkowski:1977sc,Marciano:1977wx,Cheng:1980tp,Lim:1981kv,Langacker:1988up,Alonso:2012ji}, 
\begin{align}
\text{BR}(\ell_i\to\ell_j\gamma)=\frac{\alpha_w^3 s_w^2}{256\pi^2}\Big(\frac{m_{\ell_i}}{M_W}\Big)^4\Big(\frac{m_{\ell_i}}{\Gamma_{\ell_i}}\Big)\Bigg|\frac{v_H^2}{2M^2}(Y_{\nu^c} Y_{\nu^c}^\dagger)_{ji}G_\gamma\left(\frac{M^2}{M_W^2}\right)\Bigg|^2
\label{eq:LFV-BR}
\end{align}
where $\alpha_w=g_w^2/4\pi$, $s_w^2=\sin^2\theta_w$ and $Y_{\nu^c}$ can be written in the parametrization given in Eq.~\eqref{eq:Ynu}. The loop function $G_\gamma(x)$ is given by  
\begin{align}
G_\gamma(x)=-\frac{x(2x^2+5x-1)}{4(1-x)^3}-\frac{3x^3}{2(1-x)^4}\text{log}\,x
\end{align}
Under the minimal \lfv hypothesis, i.e. that both $M$ and $\mu$ are diagonal and degenerate, we make the further assumption that the orthogonal matrix $R=\mathbb{I}$.
We show the resulting cLFV decay rates as functions of $M$ for normal ordering. 
As we already discussed, the parameter $\mu$ can be naturally small as it is determined mainly by the small parameter $\lambda_5$. 
\begin{figure}[htb!]
\centering
\includegraphics[width=0.45\textwidth]{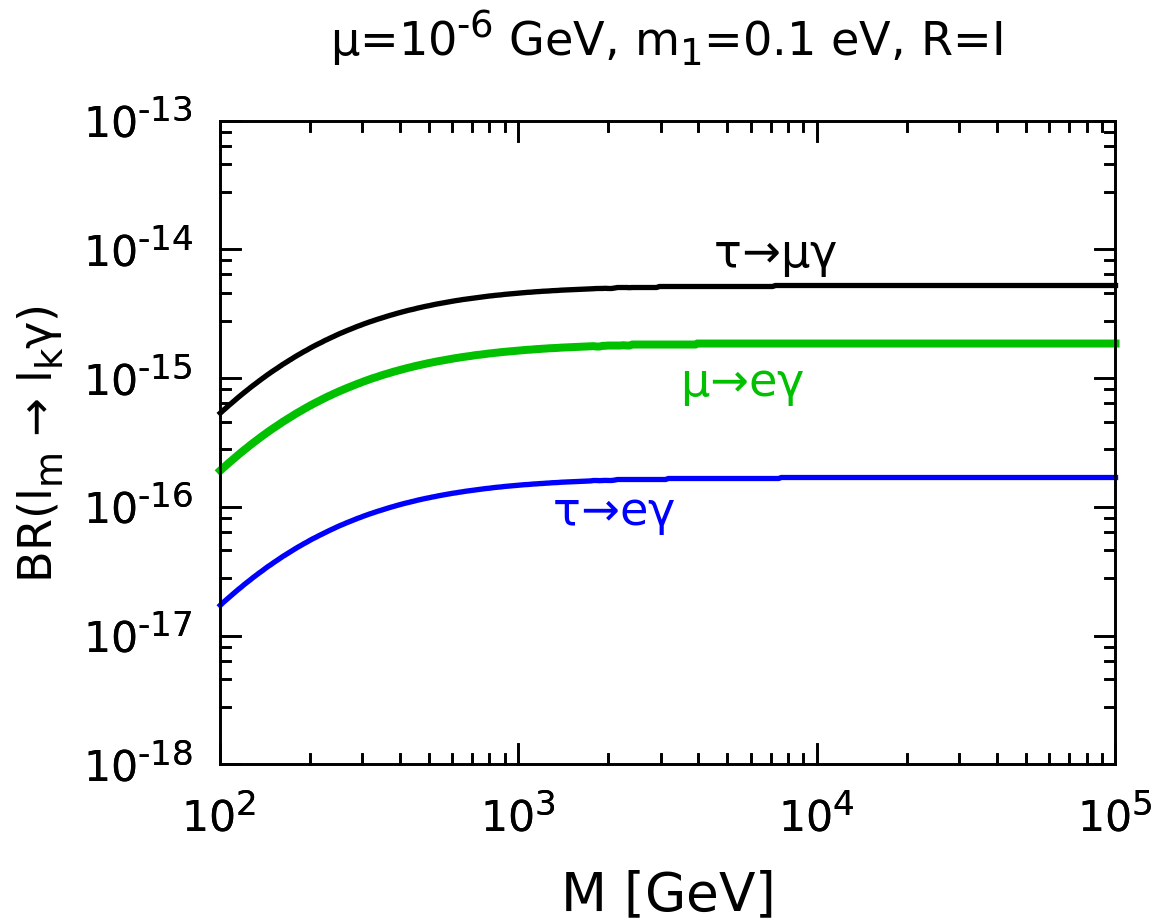}
\includegraphics[width=0.45\textwidth]{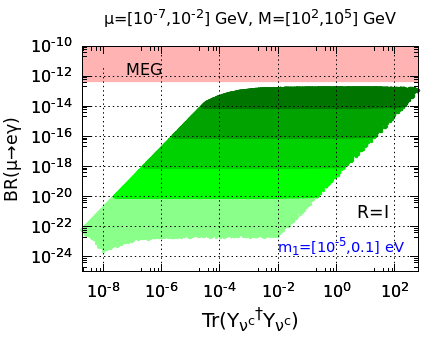}
\caption{ 
  Left panel: $\text{BR}(\mu\to e\gamma)$~(green line), $\text{BR}(\tau\to e\gamma)$~(blue line) and $\text{BR}(\tau\to \mu\gamma)$~(black line).
  The other input parameters are indicated. 
  Right panel: $\text{BR}(\mu\to e\gamma)$ as a function of $\text{Tr}(Y_{\nu^c}^{\dagger}Y_{\nu^c})$.
  We have varied the parameters as: $\mu^{\text{diag}}=\big[10^{-7}, \,10^{-2}\big]$~GeV, $M^{\text{diag}}=\big[10^{2}, \,10^{5}\big]$~GeV, $10^{-5}\,\text{eV}\leq m_1\leq 0.1\,\text{eV}$ and fixed $R=\mathbb{I}$.
  The red shaded region is excluded by the MEG experiments~\cite{MEG:2013oxv}.
  In both panels, the oscillation parameters are fixed to their best fit values~\cite{deSalas:2020pgw,10.5281/zenodo.4593330} and CP phases are set to zero.}
\label{fig:LFV}
\end{figure}

Notice that, since the value of $\mu$ can be made naturally small, the Yukawa coupling $Y_{\nu^c}$ can be large, and hence one can have large cLFV.
This is a basic feature of inverse seesaw schemes. 
This is indeed confirmed in Fig.~\ref{fig:LFV} where we assume normal ordering and choose $\mu^{\text{diag}}=10^{-6}$~GeV, $R=\mathbb{I}$ and the lightest neutrino mass $m_1=0.1$~eV. 
The green, black and blue lines shown in the left panel stand for the $\mu\to e\gamma$, $\tau\to \mu\gamma$ and $\tau\to e\gamma$ branching ratios. 
This apparent non-decoupling effect is simply an artifact of the Casas-Ibarra parametrization.
Indeed, expressed directly in terms of the Yukawa coupling $Y_{\nu^c}$ the decay rate obeys decoupling. 
In the right panel of Fig.~\ref{fig:LFV}, we show the $\text{BR}(\mu\to e\gamma)$ as a function of $\text{Tr}(Y_{\nu^c}^{\dagger}Y_{\nu^c})$. 
To obtain this we have randomly generated the diagonal entries of the matrices $M$, $\mu$ in the range
$10^2\,\text{GeV}\leq M^{\text{diag}}\leq 10^5\,\text{GeV}$, $10^{-7}\,\text{GeV}\leq \mu^{\text{diag}}\leq 10^{-2}\,\text{GeV}$, varying the lightest neutrino masses in the range
$10^{-5}\,\text{eV}\leq m_1\leq 0.1\,\text{eV}$ and expressing the Yukawa coupling $Y_{\nu^c}$ in the perturbative range using the Casas-Ibarra parametrization. 
Notice that $\mu\to e\gamma$ decay rate can be close to the present experimental bound~\cite{MEG:2013oxv} even for very small value of $\text{Tr}(Y_{\nu^c}^{\dagger}Y_{\nu^c})$.
All in all, one sees that $\mu\to e\gamma$ branching ratio can exceed current sensitivities. 
The same happens for other cLFV processes. 

\subsection{Results for the minimal ``missing partner'' $(3,2,2)$ inverse seesaw} 

\begin{figure}[htb!]
\centering
\includegraphics[width=0.45\textwidth]{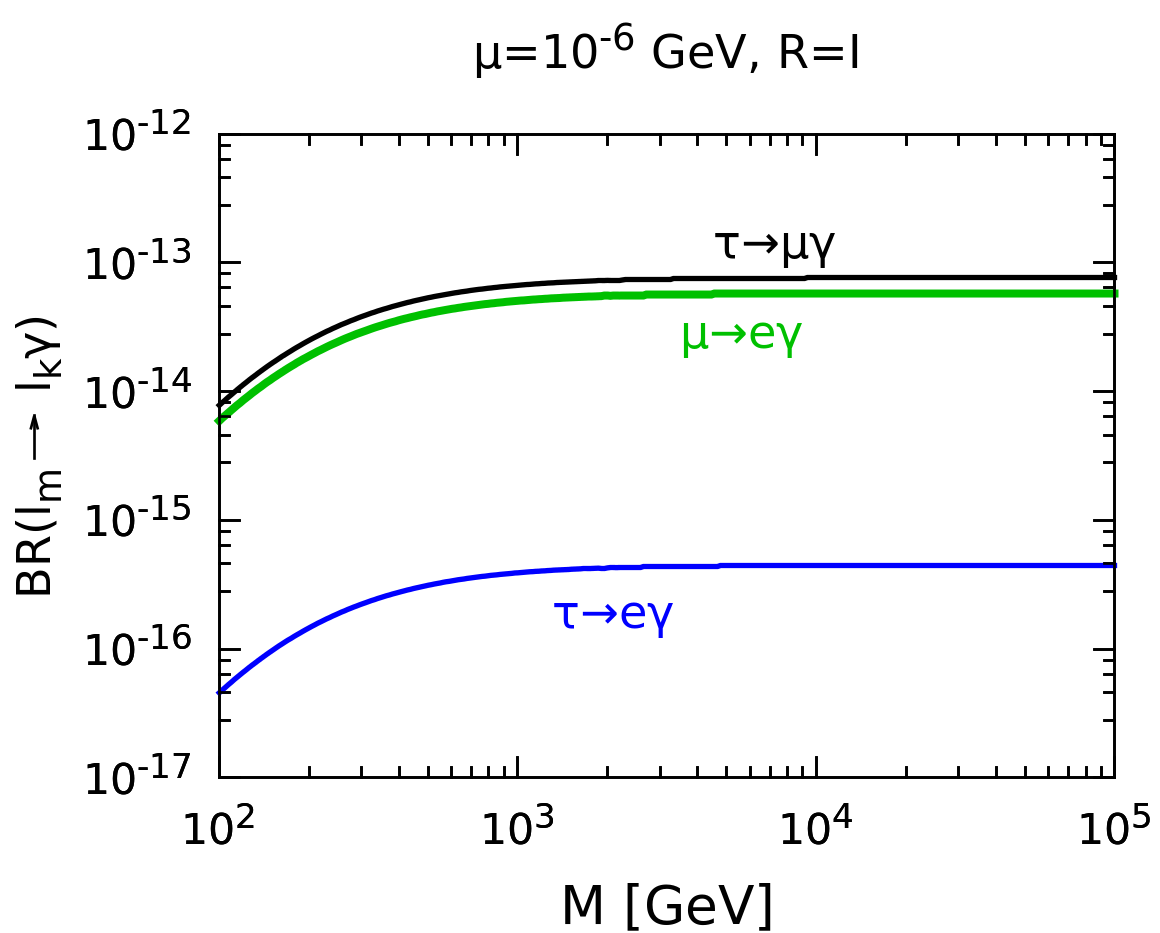}
\includegraphics[width=0.45\textwidth]{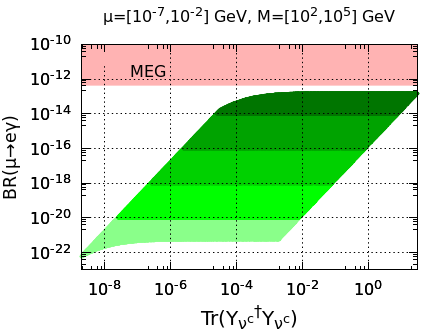}
\caption{ Same as Fig.~\ref{fig:LFV} but now for the minimal $(3,2,2)$ case with $m_1=0$.}
\label{fig:LFV2}
\end{figure}

Let us now briefly discuss the results obtained in the case of a minimal $(3,2,2)$ inverse seesaw scheme where only two copies of $\nu^c,S$ and $f$ are present.
This ensures two seesaw-induced light neutrino mass scales, with one of the three active neutrinos remaining massless, i.e. $m_1=0$.
The situation is the inverse seesaw analogue of the (3,2) scenario first discussed in~\cite{Schechter:1980gr}.
The results for such scheme are shown in Fig.~\ref{fig:LFV2}, where we adopt the same parameter choice as in Fig~\ref{fig:LFV}, in particular the same scan range for the right panel. 
Comparing the Fig.~\ref{fig:LFV} and Fig.~\ref{fig:LFV2}, one sees that one obtains analogous results.
In particular, one gets the same general enhancement of individual $\text{BR}(\ell_m\to\ell_k\gamma)$ results.
%

However, from Eq.~\eqref{eq:LFV-BR} one finds small differences, in particular, the ratios of cLFV branching ratios are modified.
Indeed, within our minimal \lfv hypothesis of diagonal and degenerate $M$ and $\mu$, one finds, 
%
\begin{align}
\frac{\text{BR}(\ell_i\to\ell_j\gamma)}{\text{BR}(\ell_k\to\ell_m\gamma)}=\frac{m_{\ell_i}^5}{m_{\ell_k}^5}\frac{\Gamma_{\ell_k}}{\Gamma_{\ell_i}}\frac{|(Y_{\nu^c}Y_{\nu^c}^{\dagger})_{ji}|^2}{|(Y_{\nu^c}Y_{\nu^c}^{\dagger})_{mk}|^2}
\label{eq:rato-LFV}
\end{align}

One sees how this double ratio depends exclusively on the ``Yukawa'' ratios $Y_{\nu^c}Y_{\nu^c}^\dagger$, which should be chosen so as to fit the measured neutrino oscillation parameters.
In Appendix~\ref{app:ratio-LFV-BR} we give the explicit results for this case. 
In such simplified scenario, taking the $3\sigma$ range of oscillation parameters from~\cite{deSalas:2020pgw,10.5281/zenodo.4593330} and neglecting CP phases, we obtain the following predictions:  
\begin{align}
&\frac{\text{BR}(\mu\to e\gamma)}{\text{BR}(\tau\to e\gamma)}\approx 31-434,\,\,\,
\frac{\text{BR}(\mu\to e\gamma)}{\text{BR}(\tau\to \mu\gamma)}\approx 0.63-0.87\,\,\,\text{and}\,\,\,
\frac{\text{BR}(\tau\to e\gamma)}{\text{BR}(\tau\to \mu\gamma)}\approx 0.002-0.022
    \end{align}
  %
\begin{figure}[htb!]
\centering
\includegraphics[width=0.45\textwidth]{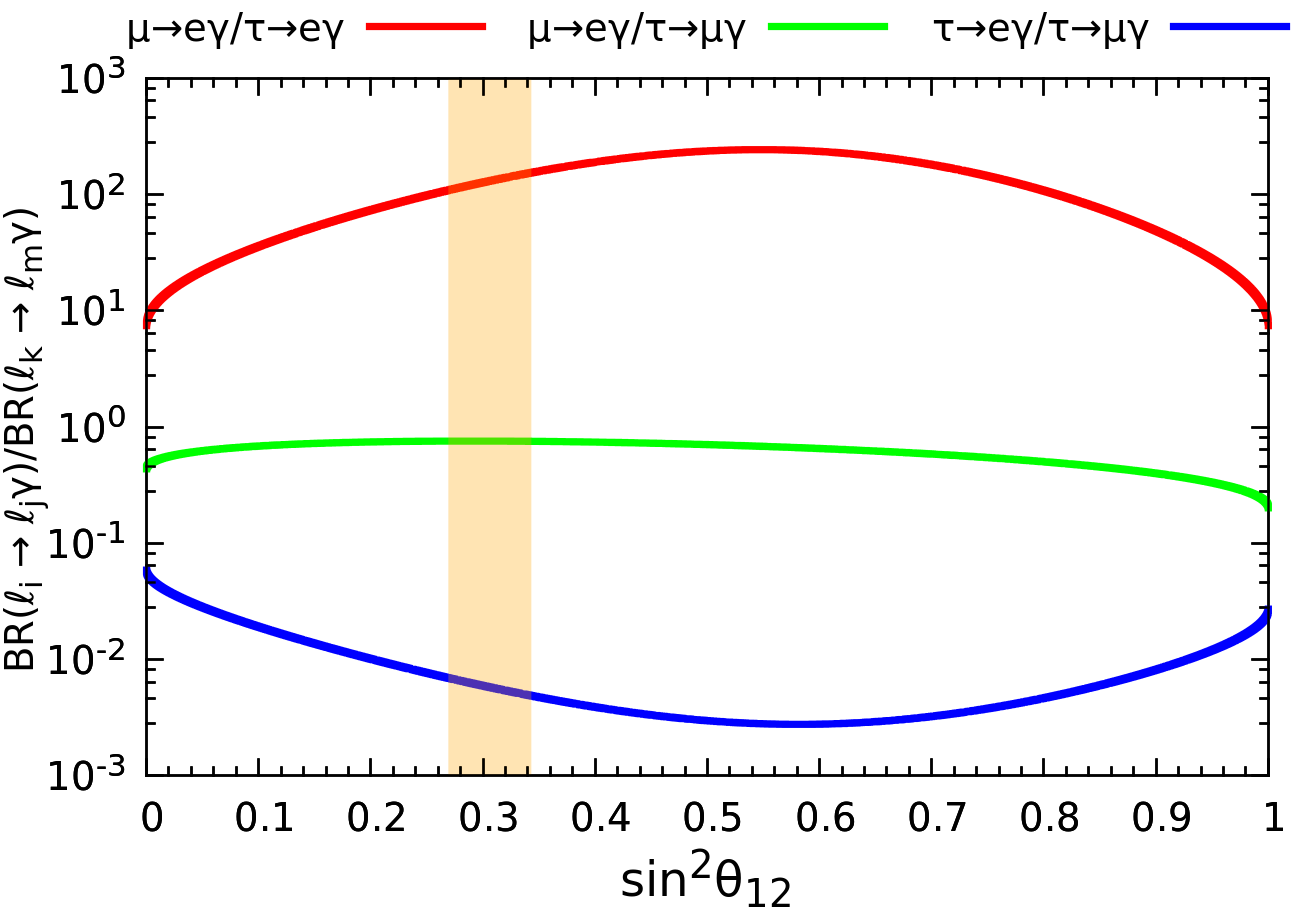}
\includegraphics[width=0.45\textwidth]{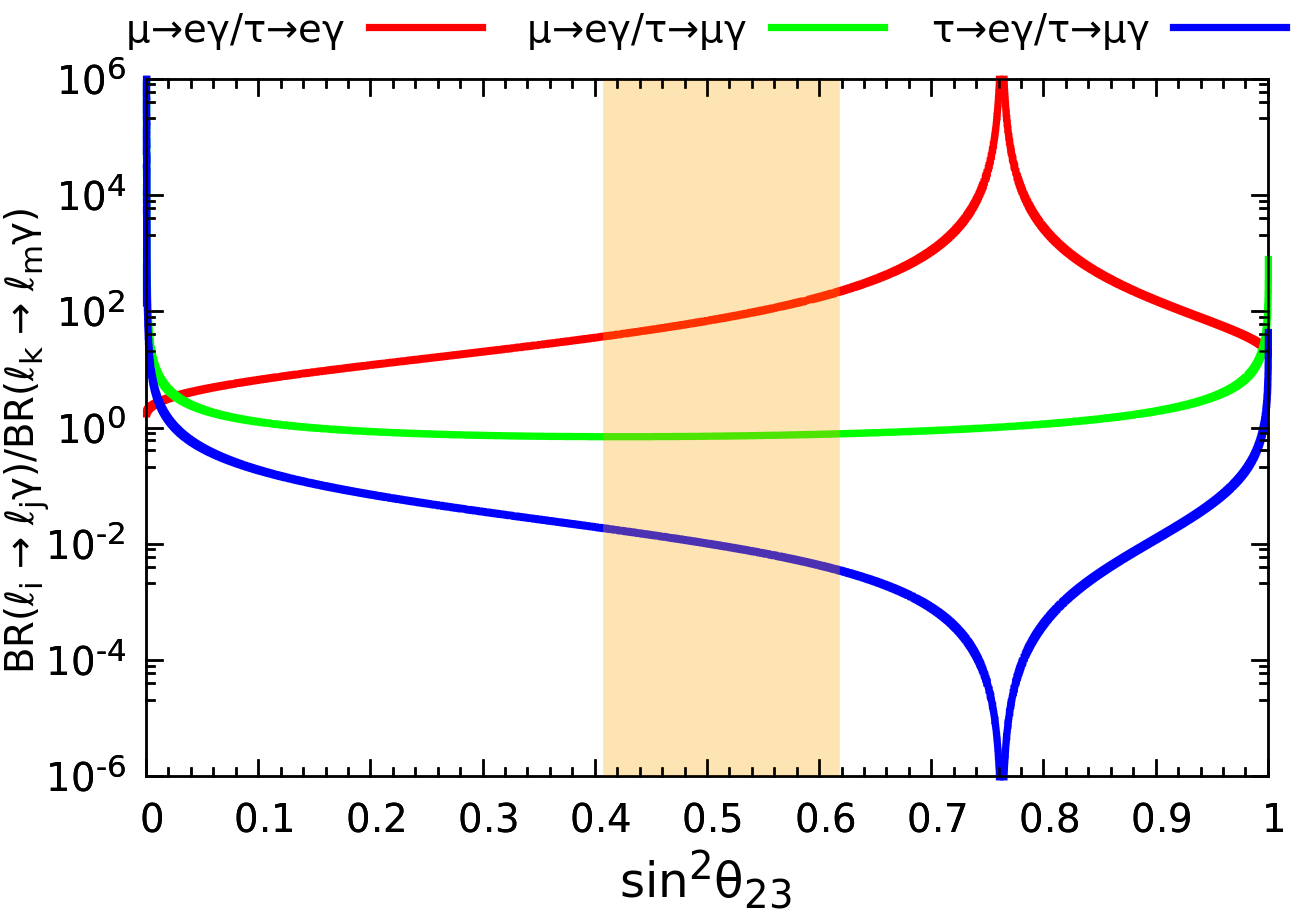}
\caption{
  $\text{BR}(\ell_i\to \ell_j\gamma)/\text{BR}(\ell_m\to \ell_k\gamma)$ versus ``solar'' and ``atmospheric'' mixing angles, $\theta_{12}$~(left) and $\theta_{23}$~(right).
  For each panel we fix the other oscillation parameter to its best fit value.
  The vertical orange band in each panel is the $3\sigma$ range from~\cite{deSalas:2020pgw,10.5281/zenodo.4593330}.}
\label{fig:theta}
\end{figure}

In Fig.~\ref{fig:theta}, we show the ratios of branching ratios $\text{BR}(\ell_i\to \ell_j\gamma)/\text{BR}(\ell_m\to \ell_k\gamma)$ %
as a function of the ``solar'' and ``atmospheric'' mixing angles, $\theta_{12}$~(left panel) and $\theta_{23}$~(right panel).
The undisplayed oscillation parameters in each case are varied randomly within their $3\sigma$ ranges~\cite{deSalas:2020pgw,10.5281/zenodo.4593330}. 
The vertical band in each panel is the $3\sigma$ range of the respective mixing angle. \\[-.4cm]

In summary, as main message we would like to stress that the rates for both lepton flavour violating~\cite{Bernabeu:1987gr},
as well as CP violating~\cite{Branco:1989bn,Rius:1989gk}, processes need not be ``neutrino-mass-suppressed''.
This point is made clearly within inverse seesaw schemes, in which cLFV processes can be non-zero even in the massless neutrino limit.
Many subsequent studies confirmed that cLFV rates can indeed be sizeable~\cite{Gonzalez-Garcia:1991brm,Ilakovac:1994kj,Deppisch:2004fa,Deppisch:2005zm,DeRomeri:2016gum}.
Moreover, as we saw above, within our minimal \emph{benchmark} where \lfv resides only in the Yukawas, the neutrino oscillation measurements get promoted into cLFV predictions.
We saw how $\mu\to e\gamma$ may be just ``around the corner'' with challenging predictions for $\tau\to\mu\gamma$ and especially $\tau\to e\gamma$.
One should note, however, that these predictions probe our minimality assumption, while the fact that cLFV can be sizeable constitutes a characteristic feature of the inverse seesaw mechanism.\\[-.4cm]

Another such feature is that the ``heavy'' isosinglet neutrinos can lie at a mass scale accessible to collider experiments.
In fact it has been suggested that these heavy neutrinos mediating neutrino mass generation could show up in the decays of the Z-boson at LEP~\cite{Dittmar:1989yg}.
Although, within the simplest \SM context, the heavy neutrino messengers have small production cross section, since they are produced only through mixing with the ordinary neutrinos,
``sitting above the Z resonance'' would help overcome this suppression~\cite{Dittmar:1989yg}. \\[-.4cm]

Subsequent studies at LEP~\cite{L3:2001zfe} and at the LHC~\cite{CMS:2018iaf} have excluded this possibility within current experimental setups. 
High-Luminosity LHC~\cite{ATLAS:2019mfr,CidVidal:2018eel}, 
may be able to provide improved search capabilities. 
In particular the use of flavor-tagging of rare LHC events of the type of $\ell_k^{\pm}\ell_m^{\mp}jj$, with $k\neq m$ may provide a powerful tool in the search
for heavy neutrino mediating ``active'' neutrino mass generation~\cite{Aguilar-Saavedra:2012dga}. 
This possibility has been considered within low-scale seesaw models containing new gauge boson portals that could enhance the Drell-Yan production cross-section~\cite{Das:2012ii}.
In fact, it was found that the LHC could reveal large rates for cLFV processes at high energies, even if the ``classic'' low-energy probes give negative results~\cite{Deppisch:2013cya}.

\section{Neutrino phenomenology}
\label{sec:neutr-phen}

A generic and important phenomenological implication of low-scale seesaw schemes, such as our inverse seesaw dark matter scenario,
is the mixing between isodoublet and isosinglet heavy neutrinos.
This implies that the lepton mixing matrix has an extended, ``rectangular'' form~\cite{Schechter:1980gr}, as it contains an admixture of the isodoublet neutrinos with heavy neutrinos.
This also leads to a non-trival nature of the neutral current~\cite{Schechter:1980gr}.
As we saw, in contrast to the Standard Model, lepton flavor is violated~\cite{Bernabeu:1987gr}, and also CP symmetry~\cite{Branco:1989bn,Rius:1989gk},
even in the limit of strictly massless (degenerate) neutrinos,
leading to a plethora of collider~\cite{Dittmar:1989yg,Gonzalez-Garcia:1990sbd}, lepton flavour and
CP violation phenomena~\cite{Gonzalez-Garcia:1991brm,Ilakovac:1994kj,Deppisch:2004fa,Deppisch:2005zm,DeRomeri:2016gum}.   \\[-.4cm]

Another implication of this scenario is that the effective mixing matrix describing the charged current leptonic weak interaction is non-unitary~\cite{Valle:1987gv,Escrihuela:2015wra,Blennow:2016jkn}.
Non-unitarity effects may not be negligible if the seesaw mechanism is realized at low scales, as in inverse seesaw. 
They could affect the prospects for probing CP violation at long-baseline neutrino experiments, since the standard CP phase could be confused with phases associated to
unitarity violation~\cite{Miranda:2016wdr,Escrihuela:2016ube}. 
Near distance experiments would be helpful in eliminating such degeneracies~\cite{Ge:2016xya,Miranda:2018yym} and could pave the way for stringent unitarity tests. 
Simulation studies should be performed to determine the CP sensitivities of upcoming oscillation experiments in the presence of unitarity violation~\cite{Fong:2017gke,Forero:2021azc}.  \\[-.2cm] 

We now turn to an interesting consequence that holds if our incomplete seesaw scenario, with only two copies of fermions $\nu^c,S$ and $f$.
In such ``missing partner''  $(3,2,2)$ inverse seesaw scotogenic scheme the effective mass parameter $m_{\beta\beta}$ characterizing the neutrinoless double beta decay amplitude
has a lower limit~\cite{Reig:2018ztc,Barreiros:2018bju,Mandal:2019ndp}.  
For heavy TeV-scale neutrino mediators, the main contribution to $m_{\beta\beta}$ comes from the light active neutrinos.
Using the symmetrical parametrization of the lepton mixing matrix $U_{\text{lep}}$~\cite{Schechter:1980gr} it can be expressed as~\cite{Rodejohann:2011vc}
%
\begin{align}
\braket{m_{\beta\beta}}\approx \Big|\sum_i U_{\text{lep},ei}^2 \,m_i\Big|=\Big|\cos^2\theta_{12}\cos^2\theta_{13}m_1+\sin^2\theta_{12}\cos^2\theta_{13}m_2 e^{2i\phi_{12}}+\sin^2\theta_{13}m_3 e^{2i\phi_{13}}\Big|.
\end{align}
%
%
\begin{figure}[htb!]
\includegraphics[width=0.65\textwidth]{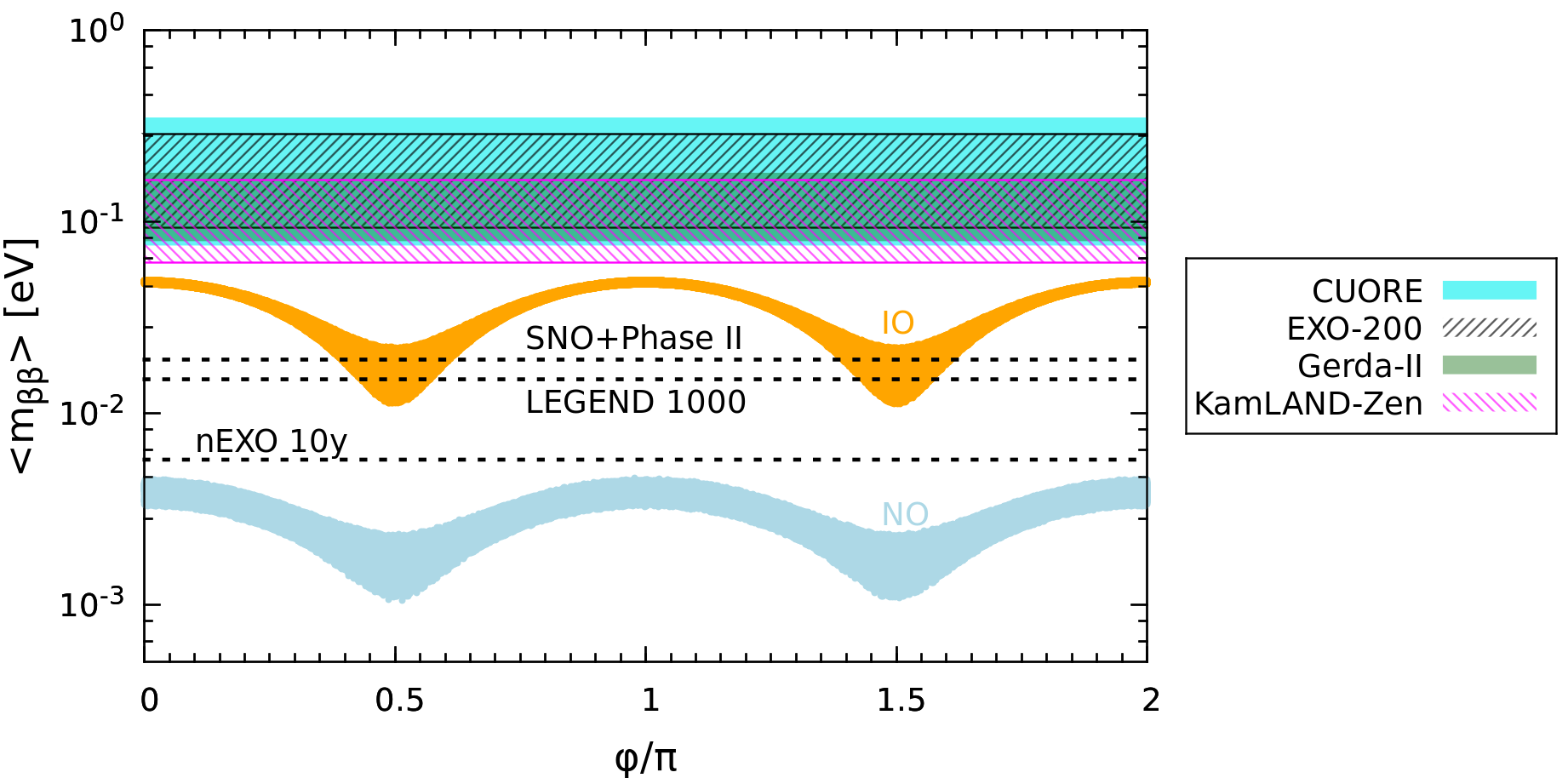}
\caption{
  Amplitude for $0\nu\beta\beta$ as a function of the majorana phase when $m_1=0$. 
  The light-blue and light-orange bands are the $3\sigma$ C.L. regions allowed by current oscillation experiments for normal and inverted mass ordering~\cite{deSalas:2020pgw,10.5281/zenodo.4593330}.
  The horizontal bands are the current limits, while the black lines show projected sensitivities. }
\label{fig:0nubb}
\end{figure}
Here $m_i$ are the light neutrino masses and $\theta_{ij}$ are the neutrino mixing angles measured in oscillation experiments. 
Since in our case the lightest neutrino is massless~(NO: $m_1 = 0$, IO: $m_3=0$), there is only one physical Majorana phase~(NO: $\phi \equiv \phi_{12}-\phi_{13}$, IO: $\phi\equiv \phi_{12}$).
In Fig.~\ref{fig:0nubb} we show the $\braket{m_{\beta\beta}}$ dependence on this phase $\phi$. 
  The lower blue band is allowed for normal ordering, whereas the orange band corresponds to the inverted ordering.
  The narrowness of the bands is due to the small allowed spread in neutrino oscillation parameters~\cite{deSalas:2020pgw,10.5281/zenodo.4593330}.  
  This should be compared with the complete case with three copies of $\nu^c$, $S$ and $f$.
  In such (3,3,3) case there can be a cancellation amongst the light neutrino amplitudes for normal ordering, so that $\braket{m_{\beta\beta}} \to 0$. 
  The horizontal bands in Fig.~\ref{fig:0nubb} show the bounds from current experiments:
  CUORE~(cyan, $\braket{m_{\beta\beta}}<0.075-0.350$~eV)~\cite{CUORE:2020ymk},
  EXO-200~(gray, $\braket{m_{\beta\beta}}< 0.093-0.286$~eV)~\cite{EXO-200:2019rkq},
  Gerda-II~(green, $\braket{m_{\beta\beta}}<0.079-0.180$~eV)~\cite{GERDA:2020xhi}
  and KamLAND-Zen~(magenta, $\braket{m_{\beta\beta}}<0.061-0.165$~eV)~\cite{KamLAND-Zen:2016pfg}.
  The three horizontal dashed black lines indicate the projected sensitivities of upcoming experiments:
  SNO+ Phase-II (0.019 eV)~\cite{SNO:2015wyx}, LEGEND-1000 (0.015 eV)~\cite{LEGEND:2017cdu} and nEXO - 10yr (0.0057 eV)~\cite{nEXO:2017nam}.
  One sees hope that upcoming experiments will be able to probe the relevant Majorana phase, at least for inverted ordering.

\section{Summary and Outlook}
\label{sec:summary}

 We have proposed that neutrino masses arise from a dark sector within the inverse seesaw mechanism.
 This way of seeding neutrino masses involves a new variant of the scotogenic scenario for radiative neutrino masses in which the ``dark'' particles are \sm singlets.
 Neutrino masses are generated when this is combined with the inverse seesaw mechanism.
  We have discussed thoroughly both the possibilities of explicit as well as dynamical lepton number violation.
  For the case of dynamical lepton number violation we have discussed in some detail the phenomenology of invisible Higgs decays with majoron emission Fig.~\ref{fig:limit-from-inv-mu},
  as well as the phenomenology of scotogenic WIMP dark matter.
  We have discussed the primordial dark matter abundance, Fig.~\ref{fig:relic}, as well as the  prospects for direct dark matter detection by nuclear recoil, Fig.~\ref{fig:fitdm}.
  We have also investigated the novel features associated to charged lepton flavour violation, Figs.~\ref{fig:LFV},\ref{fig:LFV2} and~\ref{fig:theta}.
  In contrast to conventional scotogenic models, here charged lepton flavour violation is not mediated by the exchange of dark particles.
  Instead, these processes arise \emph{a la inverse-seesaw}, due to admixture of the heavy singlet neutrinos.
  Finally, we have also briefly discussed some aspects of neutrino physics such as neutrinoless double-beta decay Fig.~\ref{fig:0nubb} and also oscillations.
  Upcoming precision neutrino oscillation experiments might reveal unitarity violation effects and/or detect the \znbb process.
  All in all, our theoretical proposal is the simplest dynamical inverse seesaw model where dark matter seeds neutrino masses radiatively, for other attempts
    see, for example~\cite{Ma:2009gu,Bazzocchi:2009kc,Baldes:2013eva,CarcamoHernandez:2018hst},
  leading to a plethora of signatures and possible experimental tests.

\begin{acknowledgments}
  This work of S.M. and J.V. is supported by the Spanish grant FPA2017-85216-P (AEI/FEDER, UE) and PROMETEO/2018/165 (Generalitat Valenciana).
  R.S. is supported by SERB, Government of India grant SRG/2020/002303.
\end{acknowledgments}
\appendix
\section{Scalar dark matter annihilation mechanisms}
\label{app:annihilation}
\begin{figure}[ht!]
\centering
\includegraphics[width=0.3\textwidth]{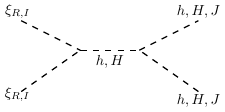}
\includegraphics[width=0.3\textwidth]{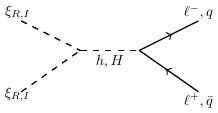}
\includegraphics[width=0.3\textwidth]{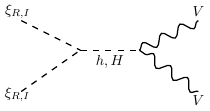}
\includegraphics[width=0.3\textwidth]{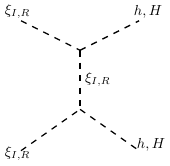}
\includegraphics[width=0.3\textwidth]{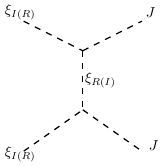}
\includegraphics[width=0.3\textwidth]{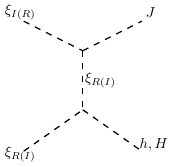}
\includegraphics[width=0.3\textwidth]{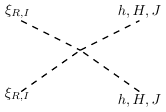}
\caption{Annihilation and coannihilation diagrams contributing to the relic abundance of $\xi_R$.}
\label{fig:annihilation-diagram}
\end{figure}

In Table.~\ref{tab:cubic-coupling} and \ref{tab:quartic-coupling}, we list the relevant cubic and quartic scalar boson couplings which play a role in dark matter analysis.
In our model the relic abundance of the lightest dark particle $\xi_R$ is determined by the annihilation and coannihilation diagrams listed in Fig.~\ref{fig:annihilation-diagram}.
\begin{table}[ht!]
\setlength\tabcolsep{0.25cm}
\centering
\begin{tabular}{| c || c |}
\hline
$\lambda_{abc}$ &  Couplings in terms of Lagrangian parameter \\
\hline
$h\xi_R\xi_R$ & $(\lambda_{\xi\sigma}+\lambda_5)v_\sigma\sin\theta + \lambda_{\Phi\xi}v_\Phi\cos\theta$ \\
$H\xi_R\xi_R$ & $(\lambda_{\xi\sigma}+\lambda_5)v_\sigma\cos\theta - \lambda_{\Phi\xi}v_\Phi\sin\theta$ \\
$h\xi_I\xi_I$ & $(\lambda_{\xi\sigma}-\lambda_5)v_\sigma\sin\theta + \lambda_{\Phi\xi}v_\Phi\cos\theta$ \\
$H\xi_I\xi_I$ & $(\lambda_{\xi\sigma}-\lambda_5)v_\sigma\cos\theta - \lambda_{\Phi\xi}v_\Phi\sin\theta$ \\
\hline
\hline
$\xi_I\xi_R J$ &  $\lambda_5 v_\sigma$  \\
\hline
\hline
$hJJ$   &  $m_h^2\sin\theta/v_\sigma$  \\
$HJJ$   &  $m_H^2\cos\theta/v_\sigma$  \\
\hline
\end{tabular}
\caption{The cubic couplings of the various scalars.}
\label{tab:cubic-coupling}
\end{table}
\begin{table}[!h]
\setlength\tabcolsep{0.25cm}
\centering
\begin{tabular}{| c || c |}
\hline
$\lambda_{abcd}$ &  Couplings in terms of Lagrangian parameter \\
\hline
$hh\xi_R\xi_R$  &   $\lambda_{\Phi\xi}\cos^2\theta + (\lambda_{\xi\sigma}+\lambda_5)\sin^2\theta$  \\
$hh\xi_I\xi_I$  &   $\lambda_{\Phi\xi}\cos^2\theta + (\lambda_{\xi\sigma}-\lambda_5)\sin^2\theta$  \\
$HH\xi_R\xi_R$  &   $\lambda_{\Phi\xi}\sin^2\theta + (\lambda_{\xi\sigma}+\lambda_5)\cos^2\theta$  \\
$HH\xi_I\xi_I$  &   $\lambda_{\Phi\xi}\sin^2\theta + (\lambda_{\xi\sigma}-\lambda_5)\cos^2\theta$  \\
\hline
\hline
$JJ\xi_R\xi_R$  &  $\lambda_{\xi\sigma}-\lambda_5$  \\
$JJ\xi_I\xi_I$  &  $\lambda_{\xi\sigma}+\lambda_5$  \\
\hline
\end{tabular}
\caption{The quartic couplings of the various scalars.}
\label{tab:quartic-coupling}
\end{table}
%
\section{Ratio of charged lepton flavour violating branching ratios}
\label{app:ratio-LFV-BR}

We now present analytical formulae for the ratios of lepton flavour violating branching ratios for the ``incomplete'' (3,2,2) inverse seesaw scheme with normal ordering.
We set the values of all CP phases to zero.

\begin{align}
&\resizebox{0.7\hsize}{!}{$\frac{\text{BR}(\mu\to e\gamma)}{\text{BR}(\tau\to e\gamma)}\approx\frac{m_\mu^5\Gamma_\tau}{m_\tau^5\Gamma_\mu}\frac{|\sin\theta_{12}(-\cos\theta_{12}\cos_{23}+\sin\theta_{12}\sin\theta_{13}\sin\theta_{23})m_2-\sin\theta_{13}\sin\theta_{23}m_3|^2}{|\sin\theta_{12}(\cos\theta_{23}\sin\theta_{12}\sin\theta_{13}+\cos\theta_{12}\sin\theta_{23})m_2-\cos\theta_{23}\sin\theta_{13}m_3|^2}$}\\
&\resizebox{0.9\hsize}{!}{$ \frac{\text{BR}(\mu\to e\gamma)}{\text{BR}(\tau\to \mu\gamma)}\approx\frac{m_\mu^5\Gamma_\tau}{m_\tau^5\Gamma_\mu}\frac{|\cos\theta_{13}(\sin\theta_{12}(\cos\theta_{12}\cos_{23}-\sin\theta_{12}\sin\theta_{13}\sin\theta_{23})m_2+\sin\theta_{13}\sin\theta_{23}m_3)|^2}{|-(\cos\theta_{23}\sin\theta_{12}\sin\theta_{13}+\cos\theta_{12}\sin\theta_{23})(\cos\theta_{12}\cos\theta_{23}-\sin\theta_{12}\sin\theta_{13}\sin\theta_{23})m_2+\cos\theta_{13}^2\cos\theta_{23}\sin\theta_{23}m_3|^2}$}\\
&\resizebox{0.9\hsize}{!}{$\frac{\text{BR}(\tau\to e\gamma)}{\text{BR}(\tau\to \mu\gamma)}\approx\frac{|\cos\theta_{13}(-\sin\theta_{12}(\cos\theta_{23}\sin\theta_{12}\sin\theta_{13}+\cos\theta_{12}\sin\theta_{23})m_2+\cos\theta_{23}\sin\theta_{13}m_3)|^2}{|-(\cos\theta_{23}\sin\theta_{12}\sin\theta_{13}+\cos\theta_{12}\sin\theta_{23})(\cos\theta_{12}\cos\theta_{23}-\sin\theta_{12}\sin\theta_{13}\sin\theta_{23})m_2+\cos\theta_{13}^2\cos\theta_{23}\sin\theta_{23}m_3|^2}$}
\end{align}
\bibliographystyle{utphys}
\bibliography{bibliography}
\end{document}